\RequirePackage{fix-cm}
\RequirePackage{rotating}

\documentclass[smallextended]{svjour3}       
\smartqed  
\usepackage{csquotes}
\usepackage{booktabs} 
\usepackage{framed}
\usepackage{graphicx}
\usepackage{amssymb}
\usepackage{tabularx}
\usepackage{graphicx}
\usepackage{url}
\usepackage{tabulary}
\usepackage{multirow}
\usepackage{enumitem}
\usepackage{hyperref}
\usepackage{hypcap}
\usepackage{lscape}
\usepackage[normalem]{ulem}
\useunder{\uline}{\ul}{}
\usepackage{longtable}
\usepackage{color}
\usepackage{hyperref}

\usepackage{booktabs}
\usepackage[flushleft]{threeparttable}
\usepackage{microtype}

\usepackage{placeins}

\usepackage{siunitx}  
\usepackage{textgreek}

\usepackage{flushend}

\usepackage{lscape}

\usepackage{rotating}

\begin{document}

\title{Predictors of Well-being and Productivity among Software Professionals during the COVID-19 Pandemic -- A Longitudinal Study}

\titlerunning{Well-being and Productivity among Software Professionals during a Pandemic}        

\author{Daniel Russo\protect\footnote{These authors contributed equally.}       \and
        Paul H. P. Hanel$^{\ast}$
        \and
        Seraphina Altnickel 
        \and
        Niels van Berkel 
}


\institute{D. Russo  \at
            Corresponding author.\\
              Aalborg University, Department of Computer Science, Aalborg, Denmark \\
              Tel.: (+45) 9940 7765\\
              \email{daniel.russo@cs.aau.dk}           
           \and
           P. H. P. Hanel \at
              University of Essex, Department of Psychology, Colchester, UK
                         \and
           S. Altnickel \at
              mia raeumerei GmbH, Berlin, Germany
                         \and
                  N. van Berkel \at
              Aalborg University, Department of Computer Science, Aalborg, Denmark
}

\date{Received: DD Month YEAR / Accepted: DD Month YEAR}

\maketitle

\begin{abstract}
The COVID-19 pandemic has forced governments worldwide to impose movement restrictions on their citizens. Although critical to reducing the virus' reproduction rate, these restrictions come with far-reaching social and economic consequences. In this paper, we investigate the impact of these restrictions on an individual level among software engineers who were working from home. Although software professionals are accustomed to working with digital tools, but not all of them remotely, in their day-to-day work, the abrupt and enforced work-from-home context has resulted in an unprecedented scenario for the software engineering community. In a two-wave longitudinal study ($N~=~192$), we covered over $50$ psychological, social, situational, and physiological factors that have previously been associated with well-being or productivity. Examples include anxiety, distractions, coping strategies, psychological and physical needs, office set-up, stress, and work motivation. This design allowed us to identify the variables that explained unique variance in well-being and productivity.   
Results include (1) the quality of social contacts predicted positively, and stress predicted an individual's well-being negatively when controlling for other variables consistently across both waves; (2) boredom and distractions predicted productivity negatively; (3) productivity was less strongly associated with all predictor variables at time two compared to time one, suggesting that software engineers adapted to the lockdown situation over time; and (4) longitudinal analyses did not provide evidence that any predictor variable causal explained variance in well-being and productivity. 
Overall, we conclude that working from home was \textit{per se} not a significant challenge for software engineers.
Finally, our study can assess the effectiveness of current work-from-home and general well-being and productivity support guidelines and provides tailored insights for software professionals.

\keywords{Pandemic \and COVID-19 \and Productivity \and Well-being \and Longitudinal Study \and Remote Work}
\end{abstract}

\section{Introduction}
\label{intro}

The mobility restrictions imposed on billions of people during the COVID-19 pandemic in the first half of 2020 successfully decreased the reproduction rate of the virus~\cite{rocklov2020covid,world2020considerations}. 
However, quarantine and isolation also come with tremendous costs on people's well-being~\cite{brooks2020} and productivity~\cite{lipsitch2020defining}.

While prior research~\cite{brooks2020} identified numerous factors either positively or negatively associated with people's well-being during disastrous events, most of this research was cross-sectional and included a limited set of predictors. Further, whether productivity is affected by disastrous events and, if so, why precisely, has not yet been investigated in a peer-reviewed article to the best of our knowledge. This is especially relevant since many companies, including tech companies, have instructed their employees to work from home~\cite{duffy2020cnn} at an unprecedented scope. Thus, it is unclear whether previous research on remote work~\cite{donnelly2015disrupted} still holds during a global pandemic while schools are closed, and professionals often have to work in non-work dedicated areas of their homes. It is particularly interesting to study the effect of quarantine on software engineers as they are often already experienced in working remotely, which might help mitigate the adverse effects of the lockdown on their well-being and productivity. Therefore, there is a compelling need for longitudinal applied research that draws on theories and findings from various scientific fields to identify variables that uniquely predict the well-being and productivity of software professionals during the first $2020$ quarantine, for both the current and potential future lockdowns. 

In the present research, we build on the literature discussed above to identify predictors of well-being and productivity. Additionally, we also include variables that were identified as relevant by other lines of research. Furthermore, we chose a different setting, sampling strategy, and research design than most of the prior literature. This is important for several reasons. 

First, many previous studies included only one or a few variables, thus masking whether other variables primarily drive the identified effects. For example, while boredom is negatively associated with well-being~\cite{farmer1986boredom}, it might be that this effect is mainly driven by loneliness, as lonely people report higher levels of boredom~\cite{farmer1986boredom}  ---  or vice versa. Only by including a range of relevant variables it is possible to identify the primary variables, which can subsequently be used to write or update guidelines to maintain one's well-being and productivity while working from home. Second, this approach simultaneously allows us to test whether models developed in an organizational context such as the two-factor theory~\cite{herzberg2017motivation} can also predict people's well-being in general and whether variables that were associated with well-being for people being quarantined also explain productivity. 
Third, while previous research on the (psychological) impact of being quarantined~\cite{brooks2020} is relevant, it is unclear whether this research is generalizable and applicable to the COVID-19 pandemic. In contrast to previous pandemics, during which only some people were quarantined or isolated, the COVID-19 pandemic strongly impacted billions globally. For example, previous research found that quarantined people were stigmatized, shunned, and rejected~\cite{lee2005experience}; this is unlikely to repeat as the majority of people are now quarantined. 
Fourth, research suggests~\cite{karesh2012ecology} that pandemics become increasingly likely due to a range of factors (\textit{e.g.}, climate change, human population growth) which make it more probable that pathogens such as viruses are transmitted to humans. This implies that it would be beneficial to prepare ourselves for future pandemics that involve lockdowns. 
Fifth, the trend to remote work has been accelerated through the COVID-19 pandemic~\cite{Meister2020}, which makes it timely to investigate which factors predict well-being and productivity while working from home. The possibility to study this under extreme conditions (\textit{i.e.}, during quarantine) is especially interesting as it allows us to include more potential stressors and distractors of productivity. This is critical. As outlined above, previous research on the advantages and challenges of remote work can presumably not be generalized to the population because mainly people from certain professions and specific living and working conditions might have chosen to work remotely. 
Sixth and finally, a longitudinal design allowed us to test for causal inferences. Specifically, in wave 1, we identified variables that explain unique variance in well-being and productivity, which we measured again in waves 2. This is important because it is possible that, for example, the amount of physical activity predicts well-being or that well-being predicts physical activity. Additionally, we are able to test whether well-being predicts productivity or vice versa  ---  previous research found that they are interrelated~\cite{krekel2019employee,carolan2017improving}.

The software engineering community has never before faced such a wide-scale lockdown and quarantine scenario during the global spread of the COVID-19 virus.
As a result, we can not build on pre-existing literature to provide tailored recommendations for software professionals.
Accordingly, in the present research, we integrate theories from the organizational~\cite{herzberg2017motivation} and psychological~\cite{masi2011meta,ryan2000self} literature, as well as findings from research on remote work~\cite{Lascau2019workers,anderson2015impact,bloom2015does} and recommendations by health~\cite{nhs_2020,sst_2020} and work~\cite{CIPD} authorities targeted at the general population, from where we derived our independent variables (or predictors). 
This longitudinal investigation provides the following contributions:

\begin{itemize}
\item First, by including a range of variables relevant to well-being and productivity, we are able to identify those variables that are uniquely associated with these two dependent variables for software professionals and thus help improve guidelines and tailor recommendations. 
\item Second, a longitudinal design allows us to explore which variables predict (rather than are predicted by) well-being and productivity of software professionals. 
\item Third, due to the current mobility restrictions imposed on billions of people we provided a unique study to understand the effects of working remotely on people's well-being and productivity. 
\end{itemize}

Our results are relevant to the software community because the number of knowledge workers who are at least partly working remotely is increasing~\cite{gallup2020}, yet the impact of working remotely on people's health and productivity is not well understood yet~\cite{mann2003psychological}. 
So far, we have only evidence regarding to the working activity distribution of developers working from home during the lockdown, compared to a typical office day, which seems to be the same~\cite{russo2021daily}.
We focus on well-being and productivity as dependent variables because both are crucial for our way of living. According to the Universal Declaration of Human Rights, well-being is a fundamental human right, and productivity allows us to maintain a certain standard of living and affect our overall well-being.
For this reason, we investigated which are the most relevant factors associated with our two dependent variables.
To do so, we started with those factors suggested by the literature (\textit{e.g.}, boredom, anxiety, routines) and validated those associations through multiple statistical analyses~\cite{russo2019soft}.
Thus, our research question is:
\\[.1in]
\noindent\textit{\textbf{Research Question}}: \textit{
What are the relevant predictors of well-being and productivity for software engineers working remotely during a pandemic?
}
\\[.1in]

In the remainder of this paper, we describe the related work about well-being in quarantine and productivity in remote work in Section~\ref{sec:related}, followed by a discussion about the research design of this longitudinal study in  Section~\ref{sec:design}. 
The analysis is described in Section~\ref{sec:analysis}, and results are discussed in Section~\ref{sec:results}.
Implications and recommendations for software engineers, companies, and any remote-work interested parties is then outlined in Section~\ref{sec:discussion}.
Finally, we conclude this paper by outlying future research directions in Section~\ref{sec:conclusion}.

\section{Related Work}
\label{sec:related}

\subsection{Well-Being in Quarantine}
To slow down the spread of pandemics, it is often necessary to quarantine a large number of people~\cite{rocklov2020covid,world2020considerations} and enforce social distancing to limit the spread of the infection~\cite{anderson2020will}. 
This typically implies that only people working in essential professions such as healthcare, police, pharmacies, or food chains, such as supermarkets, are allowed to leave their homes for work. 
If possible, people are asked to work remotely from home. However, such measures are perceived as drastic and can have severe consequences on people's well-being~\cite{brooks2020,lunn2020using}. 

Previous research has found that being quarantined can lead to anger, depression, emotional exhaustion, fear of infecting others or getting infected, insomnia, irritability, loneliness, low mood, post-traumatic stress disorders, and stress~\cite{sprang2013posttraumatic,hawryluck2004sars,lee2005experience,marjanovic2007relevance,reynolds2008understanding,bai2004survey}. 
The fear of getting infected and infecting others, in turn, can become a substantial psychological burden~\cite{kim2015public,prati2011social}. 
Also, a lack of necessary supplies such as food or water~\cite{wilken2017knowledge} and insufficient information from public health authorities adds on to increased stress levels~\cite{caleo2018factors}. 
The severity of the psychological symptoms correlated positively with the duration of being quarantined and symptoms can still appear years after quarantine has ended~\cite{brooks2020}. 
This makes it essential to understand what differentiates those whose mental health is more negatively affected by being quarantined from those who are less strongly affected. However, a recent review found that no demographic variable was conclusive in predicting whether someone would develop psychological issues while being quarantined~\cite{brooks2020}. Moreover, prior studies investigating such predictors focused solely on demographic factors (\textit{e.g.}, age or number of children~\cite{hawryluck2004sars,taylor2008factors}). 
This suggests that additional research is needed to identify psychological and demographic predictors of well-being. For example, prior research suggested that a lack of autonomy, which is an innate psychological need~\cite{ryan2000self}, negatively affects people's well-being and motivation~\cite{calvo2020health}, yet evidence to support this claim in the context of a quarantine is missing. 

To ease the intense pressure on people while being quarantined or in isolation, research and guidelines from health authorities provide a range of solutions on how an individual's well-being can be improved. 
Some of these factors lie outside of the control for individuals, such as the duration of the quarantine, or the information provided by public authorities~\cite{brooks2020}. 
In this study, we therefor focus on those factors that are within the control of individuals. 
However, investigating such factors independently might make little sense since they are interlinked. 
For example, studying the relations between anxiety and stress with well-being in isolation is less informative, as both anxiety and stress are negatively associated with well-being~\cite{de2014emotion,Spitzer2006GAD7}. However, knowing which of the two has a more substantial impact on people's well-being above and beyond the other is crucial, as it allows \textit{inter alia} policymakers, employers, and mental health support organizations to provide more targeted information, create programs that are aimed to reduce people's anxiety or stress levels, and improve people's well-being, since anxiety and stress are conceptually independent constructs. For example, stress has usually a more specific cause, is temporary, and easier to treat (\textit{e.g.}, by working less). In contrast, anxiety is more unspecific, longer-lasting, and can require professional attention~\cite{johnston2020stress}.
Thus, it is essential to study these variables together rather than separately.

\subsection{Productivity in Remote Work}
The containment measures not only come at a cost for people's well-being but they also negatively impact their productivity. For example, the International Monetary Fund (IMF) estimated in October $2020$ that the World GDP would drop by $4.4$\% as a result of the containment measures taken to reduce the spread of COVID-19 -- with countries particularly hit by the virus, such as Italy, would experience a drop of over $10$\%~\cite{imf2020world}. This expected drop in GDP would be significantly larger if many people were unable to work remotely from home. However, previous research on the impact of quarantine typically focused on people's mental and physiological health, thus providing little evidence on the effect on productivity of those who are still working. Luckily, the literature on remote work, also known as telework, allows us to get a broad understanding of the factors that improve and hinder people's productivity during quarantine. \par
The number of people working remotely has been growing in most countries already before the COVID-19 pandemic~\cite{owl_labs_2019,gallup2020}. Of those working remotely, $57$\% do so for all of their working time. The vast majority of remote workers, $97$\% would recommend others to do the same~\cite{buffer2020}, suggesting that the advantages of remote work outweigh the disadvantages. The majority of people who work remotely do so from the location of their home~\cite{buffer2020}.

Working remotely has been associated with a better work-life balance, increased creativity, positive affect, higher productivity, reduced stress, and fewer carbon emissions because remote workers commute less~\cite{owl_labs_2019,buffer2020,anderson2015impact,bloom2015does,vega2015within,baruch2000teleworking,cascio2000managing}. However, working remotely also comes with its challenges. For example, challenges faced by remote workers include collaboration and communication (named by $20$\% of $3,500$ surveyed remote workers), loneliness ($20$\%), not being able to unplug after work (18\%), distractions at home ($12$\%), and staying motivated ($7$\%)~\cite{buffer2020}. While these findings are informative, it is unclear whether they can be generalized. For instance, if mainly those with a long commute or those who feel comfortable working from home might prefer to work remotely, it would not be possible to generalize to the general working population. 

A pandemic such as the one caused by COVID-19 in $2020$ forces many people to work remotely from home. Being in a frameless and previously unknown work situation without preparation intensifies common difficulties in remote work. Adapting to the new environment itself and dealing with additional challenges adds on to the difficulties already previously identified and experienced by remote workers, and could intensify an individual's stress and anxiety and negatively affect their working ability. The advantages of remote work might, therefore, be reduced or even be reversed. Substantial research is needed to understand further what enables people to work effectively from home while being quarantined~\cite{kotera2020psychological}. The current situation shows how important research in this field is already. Forecasts indicate that remote work will grow on an even larger scale than it did over the past years~\cite{owl_labs_2019,gallup2020}, therefore research results on predictors of productivity while working remotely will increase in importance. Some guidelines have been developed to improve people's productivity, such as the guidelines proposed by the Chartered Institute of Personnel and Development, an association of human resource management experts~\cite{CIPD}. Examples include designating a specific work area, wearing working clothes, asking for support when needed, and taking breaks. However, while potentially intuitive, empirical support for those particular recommendations is still missing. 

Adding to the complexity, the measurement of productivity, especially in software engineering, is a debated issue, with some authors suggesting not to consider it at all~\cite{Ko2019}.
Nevertheless, individual developer's productivity has a long investigation tradition~\cite{sackman1968exploratory}.   
Prior work on developer productivity primarily focused on developing software tools to improve professionals' productivity~\cite{kersten2006using} or identifying the most relevant predictors, such as task-specific measurements and years of experience~\cite{dieste2017empirical}. 
Similarly, understanding relevant skillsets of developers that are relevant for productivity has also been a typical line of research~\cite{li2015makes}. 
Eventually, as La Toza et al. pointed out, measuring productivity in software engineering is not just about using tools; instead, it is about how they are used and what is measured~\cite{latoza2020explicit}.

\section{Research Design}
\label{sec:design}

There are dozens of definitions and operationalizations of well-being~\cite{linton2016review}. In the present research, we adopt a common broad and global definition of subjective well-being, following Diener~\cite{diener_beyond_2009} who defined well-being as ``the fact that the person subjectively believes his or her life is desirable, pleasant, and good" (p. 1). In other words, well-being can be understood as whether a person is overall satisfied with their lives and believes the conditions of their lives are  excellent~\cite{diener1985satisfaction}. Psychological variables such as anxiety, loneliness, or stress can be understood as parts of general well-being or as determinants thereof~\cite{keyes2003dimensions}. We consider those variables as determinants and assess the degree with which variables play a role in software engineers' overall well-being.

\begin{figure}[h!]
    \centering
    \includegraphics[width=.9\textwidth]{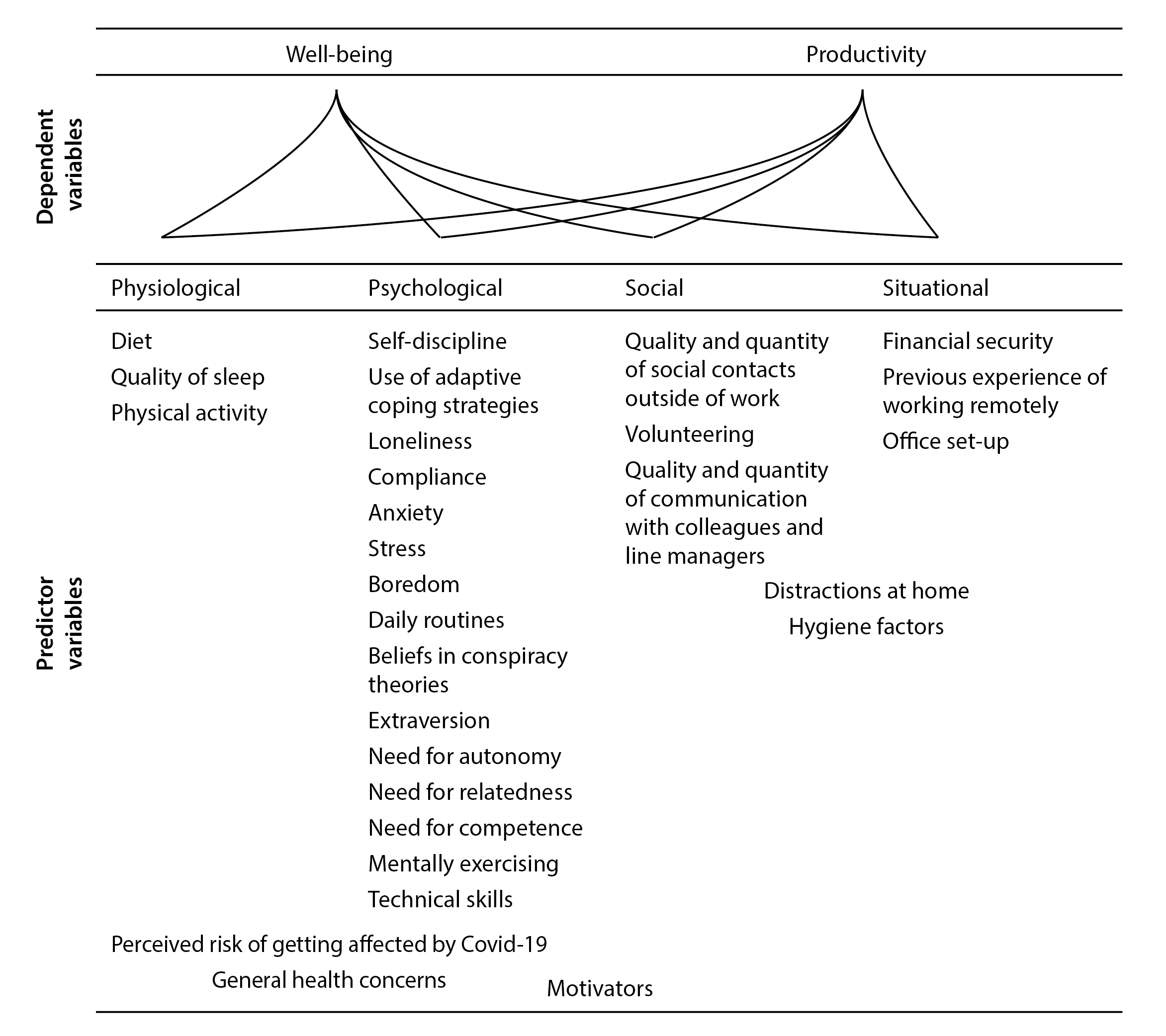}
    \caption{Overview of the independent and dependent variables}
    \label{fig:model}
\end{figure}

The variables we plan to measure in the present two-wave longitudinal study are displayed in Figure~\ref{fig:model}. To facilitate its interpretation, we categorized the variables into four broad sets of predictors, partly overlapping. To summarize, while the initial selection of predictors is theory-driven, based on previous research, or recent guidelines, the selection of predictors included in the second wave is data-driven. In other words, we used a two-step approach to select our variables: First, the initial selection of 51 predictors as based on existing theory, which we then reduced based on how strongly they are associated with well-being and productivity for an initial multiple regression analysis and the subsequent longitudinal analysis. This approach helped us to focus on the most relevant predictors while keeping their amount manageable.

During the COVID-19 pandemic, many governments and organizations have called for volunteers to support self-isolation (see, for example, ~\cite{nhs_2020support,nyc_2020}). While also relevant to the community at large, research suggests that acts of kindness positively affect people's well-being~\cite{buchanan2010acts}. Additionally, volunteering has the benefit of leaving one's home for a legitimate reason and reducing cabin fever. We, therefore, decided to include volunteering as a potential predictor for well-being. \par
Coping strategies such as making plans or reappraising the situation are, in general, effective for one's well-being~\cite{webb2012dealing,carver1989assessing}. For example, altruistic acceptance  ---  accepting restrictions because it is serving a greater good  ---  while being quarantined was negatively associated with depression rates three years later~\cite{liu2012depression}. Conversely, believing that the quarantine measures are redundant because COVID-19 is nothing but ordinary flu or was intentionally released by the Chinese government (\textit{i.e.}, beliefs in conspiracy theories) will likely lead to dissatisfaction because of greater feelings of non-autonomy. Indeed, beliefs in conspiracy theories are associated with lower well-being~\cite{freeman2017concomitants}. \par 
We further propose that three needs are relevant to people's well-being and productivity~\cite{ryan2000self}. Specifically, we propose that the need for autonomy and competence are deprived of many people who are quarantined, which negatively affects well-being and motivation~\cite{calvo2020health}. Further, we propose that the need for competence was deprived, mainly for the people who cannot maintain their productivity-level. This might especially be the case for those living with their families. In contrast, the need for relatedness might be over satisfied for those living with their family. \par 
Another important factor associated with one's well-being is the quality of one's social relationships~\cite{birditt2007relationship}. As people have fewer opportunities to engage with others they know less well, such as colleagues in the office or their sports teammates, the quality of existing relationships becomes more important, as having more good friends facilitates social interactions either in person (\textit{e.g.}, with their partner in the same household) or online (\textit{e.g.}, video chats with friends).  \par
Moreover, we expect that extraversion is linked to well-being and productivity. For example, extraverted people prefer more sensory input than introverted people~\cite{ludvigh1974extraversion}, which is why they might struggle more with being quarantined. Extraversion correlated negatively with support for social distancing measures~\cite{carvalho2020personality}, which is a proxy of stimulation (\textit{e.g.}, being closer to other people, will more likely result in sensory stimulation). 
Finally, research on productivity predictors while working from home can be theoretically grounded in models of job satisfaction and productivity, such as Herzberg's two-factor theory~\cite{herzberg2017motivation}. This theory states that causes of job satisfaction can be clustered in motivators and hygiene factors. Motivators are intrinsic and include advancement, recognition, work itself, growth, and responsibilities. Hygiene factors are extrinsic and include the relationship with peers and supervisor, supervision, policy and administration, salary, working conditions, status, personal link, and job security. Both factors are positively associated with productivity~\cite{bassett2005does}. As there are little differences between remote and on-site workers in terms of motivators and hygiene factors~\cite{green2009exploring}, the two-factor theory provides a good theoretical predictor of productivity of people working remotely.

\subsection{Participants}

Our two-wave study covers an extensive set of $51$ predictors, as identified above. 
Based on the literature mentioned earlier, we expected the strength of the association between the predictors and the outcomes' well-being and productivity to vary between medium to large. 
Therefore, we assumed for our power analysis a medium-to-large effect size of $f^2~=~.20$ and a power of $.80$. 
Power analysis with G*Power 3.1.9.4~\cite{faul2009statistical} revealed that we would need a sample size of $190$ participants. 

To collect our responses, we used Prolific,\footnote{www.prolific.co} a data collection platform, commonly used in Computer Science (see \textit{e.g.},~\cite{Hosio2020CrowdsourcingDiets}). 
We opted for this solution because of the high reliability, replicability, and data quality of dedicated platforms, especially as compared with \textit{e.g.} mailing lists~\cite{peer2017beyond,palan2018prolific}.

Specifically, the use of crowdsourcing platforms allows us to (i) avoid overloading members of mailing lists or groups on social media (\textit{e.g.}, LinkedIn, Discord) with unsolicited participation requests; (ii) recruit participants of the target population (\textit{e.g.}, only software engineers) using automatic screening option, or by running \textit{ad hoc} screening studies; (iii) recruit only participants who are interested in the research; (iv) have a high degree of control with regards to data quality since participants can get reputed without paying them and lowering their acceptance rate, which will influence future recruitment; (v) compensate participants for their time so that they will take care of the responses due to a contractual obligation; and (vi) minimize self-selection bias, since potential candidates are randomly assigned to each study (if they meet the inclusion criteria), lowering the probability that opinionated individuals take part to the survey. In sum, it is a convenient, fair, and efficient way to recruit survey informants \cite{bethlehem2010selection}. For these reasons, crowdsourcing platforms are commonly used in studies published in top-tier outlets \cite{anumanchipalli2019nature,kraft2018nature,berens2020nature}.

To administer the surveys, we used Qualtrics\footnote{www.qualtrics.com} and shared it on the Prolific platform. In order to ensure data quality and consistency, and to account for potential dropout of participants between the two waves, we invited almost $500$ participants who were identified as software engineers in a previous study~\cite{russo2020gender} to participate in a screening study in April $2020$.
The $483$ candidates already passed a multi-stage screening process, as described by Russo \& Stol, to ensure the highest possible data quality through cluster sampling~\cite{baltes2020sampling}.

To run a coherent and reliable investigation, we only recruited software engineers who were living similar experiences both from a professional and personal perspective (i.e., working remotely during a lockdown). Thus, we performed a screening study completed by $305$ software professionals who agreed to participate in a multi-wave study.
From the $305$ candidates, we excluded those living in countries with unclear, mixed policies or early reopening (\textit{e.g.}, Denmark, Germany, Sweden) and professionals working from home during the lockdown less than $20$h a week (\textit{i.e.}, excluding unemployed, or developers which had to work in their offices).
In both waves, all participants stated that they were working from home during the lockdown (a negative answer of one of these two conditions would have resulted in discarding the delivered responses from our data set).

As a result of this screening, in the first wave of data collection, which took place in the week of April $20$ - $26$ $2020$, $192$ participants completed the first survey.
Participation in the second wave (May $4$ - $10$) was high ($96$\%), with $184$ completed surveys.
Participants have been uniquely identified through their Prolific ID, which was essential to run the longitudinal analysis while allowing participants to remain anonymous.

Additionally, to enhance our responses' reliability, in each survey we included three test items (\textit{e.g.}, ``Please select response option `slightly disagree'''). 
As none of our participants failed at least two of the three test items, all participants reported working remotely and answered the survey in an appropriate time frame, and we did not exclude anyone.

The $192$ participants' mean age was $36.65$ years ($SD~=~10.77$, range$=19–63$; $154$ men, $38$ women). Participants were compensated in line with the current US minimum wage (average completion time $1202$ seconds, $SD=795.41$).
Out of our sample of $192$ participants, $63$ were based in the UK, $52$ were based in the USA, $19$ from Portugal, $10$ from Poland, $7$ from Italy, $6$ from Canada, and the remaining $35$ participants from other countries in Europe. A minority of $30$ participants reported living alone, with most participants ($162$) reported living together with others -- including babies, children, and adults. Our participants are employed primarily at private companies ($156$), followed by $30$ participants employed at a public institution. Six participants indicated to work either for a different type of company or were unsure how to categorize their employer. When asked in our screening study what percentage of their time participants were working remotely (i.e., not physically in their office) over the past 12 months, $54.7\%$ reported $25\%$ or less of their time, $15.6\%$ between $25\%$ -- $50\%$, $2.1\%$ between $50\%$ -- $75\%$, and $27.1\%$ of the participants to work remotely for at least $75\%$ of their time.

\subsection{Longitudinal design} 
We employed a longitudinal design, with two waves set two-weeks apart from each other towards the end of the lockdown, which allowed us to test for internal replication. 
Also, running this study towards the end of the lockdowns in the vast majority of countries allowed participants to provide a more reliable interpretation of lockdown conditions. We chose a period of two weeks because we wanted to balance change in our variables over time with the end of a stricter lockdown that was discussed across many countries when we run wave 2. Many of our variables are thought to be stable over time. That is, a person's scores on X at time 1 is strongly predictive of a person's scores on X at time 2 (indeed, the test-retest reliabilities we found support this assumption, see Table \ref{tab:correlations}). The closer the temporal distance between wave 1 and 2, the higher the stability of a variable. In other words, if we had measured the same variables again after only one or two days, there would not have been much variance that could have been explained by any other variable, because X measured at time 1 already explains almost all variance of X measured at time 2. In contrast, we aimed to collect data for wave 2 while people were still quarantined. If at time 1 of the data collection people would still be in lockdown and at time 2 the lockdown would have been eased, this would have included a major confounding factor. Thus, to balance those two conflicting design requirements, we opted for a two weeks break in between the two waves.

We describe the measures of the two dependent (or outcome) variables in Subsection \ref{ssec:dependent}. 
Predictors (or independent variables) are explained in Subsections \ref{ssec:independent1}, \ref{ssec:Physiological}, \ref{ssec:independent2}, and \ref{ssec:independent3}.
Wherever possible, we relied on validated scales. 
If this was not possible (\textit{e.g.}, COVID-19 specific conspiracy beliefs), we created a scale.
In those cases, we followed scale development guidelines, including avoiding negatives and especially double-negatives, two-statements within one item, and less common expressions~\cite{boateng2018best}.
The questionnaires are reported in the Supplemental Materials, while the summary of the measurement instruments with their readabilities are listed in Table \ref{tab:constructs}.
Test score reliability has been measured using Cronbach's alpha and reported for each instrument. If the instrument was used in wave 1 and wave 2, we report both Cronbach's alpha values (i.e., $\alpha_{time 1}$, $\alpha_{time 2}$); if we used it only in the first wave, we reported only the result for wave 1 ($\alpha_1$))
Additionally, we also explore whether there are any mean changes in the variables we measured at both times (\textit{e.g.}, has people's well-being changed?), and mean differences between gender and people based on different countries.

\subsection{Measurement of the dependent variables} 
\label{ssec:dependent}

\textbf{Well-being} was measured with an adapted version of the 5-item Satisfaction with Life Scale~\cite{diener1985satisfaction}. We adapted the items to measure satisfaction with life in the past week, which is in line with recommendations that the scale can be adapted to different time frames~\cite{pavot2009review}. Example items include ``The conditions of my life in the past week were excellent'' and ``I was satisfied with my life in the past week''. Responses were given on a $7$-point Likert scale ranging from $1$ (Strongly disagree) to $7$ (Strongly agree, $\alpha_{time 1}~=~.90$, $\alpha_{time 2}~=~.90$). \par
\textbf{Productivity} was measured relative to the expected productivity. We contrasted productivity in the past week with the participant's expected productivity (\textit{i.e.}, productivity level without the lockdown). As we recruited participants working in different positions, including freelancers, we can neither use objective measures of productivity nor supervisor assessments and rely on self-reports. We expect limited effects of socially desirable responses as the survey was anonymous. 
We operationalized productivity as a function of time spent working and efficiency per hour, compared to a normal week. Specifically, we asked participants: ``How many hours have you been working approximately in the past week?'' (Item P1) and ``How many hours were you expecting to work over the past week assuming there would be no global pandemic and lockdown?'' (Item P2). Finally, to measure perceived efficiency, we asked: ``If you rate your productivity (\textit{i.e.}, outcome) per hour, has it been more or less over the past week compared to a normal week?'' (Item P3). Responses to the last item were given on a bipolar slider measure ranging from `$100\%$ less productive' to `0\%: as productive as normal' to `$\geq 100\%$ more productive' (coded as -$100$, $0$, and $100$). To compute an overall score of productivity for each participant, we used the following formula: productivity~=~(P1/P2) $\times$ ((P3 + $100$)/$100$). Values between $0$ and $.99$ would reflect that people were less productive than normal, and values above $1$ would indicate that they were more productive than usual. For example, if one person worked only $50$\% of their normal time in the past week but would be twice as efficient, the total productivity was considered the same compared to a normal week. \par
We preferred this approach over the use of other self-report instruments, such as the WHO's Health at Work Performance Questionnaire~\cite{kessler2003world}, because we were interested in the change of productivity while being quarantined as compared to `normal' conditions. The WHO's questionnaire, for example, assesses productivity also in comparison to other workers. We deemed this unfit for our purpose as it is unclear to what extent software engineers who work remotely are aware of other workers' productivity. Also, our measure consists of only three items and showed good test-retest reliability (Table~\ref{tab:correlations}). Test-retest reliability is the agreement or stability of a measure across two or more time-points. A coefficient of 0 would indicate that responses at time 1 would not be linearly associated with those at time 2, which is typically undesired. Higher coefficients are an additional indicator of the reliability of the measures, although they can be influenced by a range of factors such as the internal consistency of the measure itself and external factors. For example, the test-retest reliability for productivity is $r~=~.50$ lower than for most other variables such as needs or well-being, but this is because the latter constructs are operationalized as stable over time. In contrast, productivity can vary more extensively due to external factors such as the number of projects or the reliability of one's internet connection.\footnote{We measured productivity differently from well-being and the other psychological variables. Productivity was measured as a change score whereas well-being, for example, as how people felt over the past week. Measuring productivity in the same way as well-being -- \textit{e.g.}, ``how productive have you been in the past week'' -- would have been confounded with likely different reference groups (\textit{e.g.}, are those working part-time comparing their productivity with someone working full-time or the odd colleague who has been working 100h in the past week?). To address this methodological limitation, we also measured productivity in a more comparable way to the other constructs by asking ``How many tasks that you were supposed to complete last week did you effectively manage to complete?'' Responses were given on a slider measure ranging from $0$ to $100\%$. This item correlated with $r = .34, p < .001$ with our main productivity measure (the change-score measure, an additional reliability evaluation which aims to remove the measurement error from the two observed measures \cite{oakes2001statistical}), further supporting the reliability of our measure. More importantly, however, the correlations of the task-completed item with the other variables were very similar to the change-score measure. For example, change-score productivity correlated at $r = .18, p = .01$ with well-being (Table~\ref{tab:correlations}), whereas the task-completed item correlated also positively with well-being, $r = .15, p = .04$. Other correlations with variables we discuss in more detail below of the change-score and the task-completed measured of productivity were $r = -.33$ and $-.25$ with boredom, $r = .37$ and $.38$ with need for competence, $r = .30$ and $.24$ with quality and quantity of communication, and with $r = -.34$ and $-.39$ with distractions (all $p$s $< .001$). Together, this suggests that both ways of measuring productivity are reliable.}

\subsection{Psychological factors} 
\label{ssec:independent1}
\textbf{Self-discipline} was measured with $3$-items of the Brief Self-Control Scale~\cite{tangney2004high}. Example items include ``I am good at resisting temptation'' and ``I wish I had more self-discipline'' (recoded). Responses were registered on a $5$-point scale ranging from $1$ (Not at all) to $5$ (Very; $\alpha~=~.64$). \par
\textbf{Coping strategies} were measured using the $28$-item Brief COPE scale, which measures $14$ coping dimensions~\cite{Carver1997BriefCOPE}. Example items include ``I've been trying to come up with a strategy about what to do'' (Planning) and ``I've been making fun of the situation'' (Humor). Responses were on a $5$-point scale ranging from $0$ (I have not been doing this at all) to $4$ (I have been doing this a lot). The internal consistencies were satisfactory to very good for two-item scales: Self-distraction ($\alpha~=~.65$), active coping ($\alpha~=~.61$), Denial ($\alpha~=~.66$), Substance use ($\alpha~=~.96$), Use of emotional support ($\alpha~=~.77$), Use of instrumental support ($\alpha~=~.75$), Behavioral disengagement ($\alpha_1~=~.76$, $\alpha_2~=~.71$), Venting ($\alpha~=~.65$), Positive reframing ($\alpha~=~.72$), Planning ($\alpha~=~.76$), Humor ($\alpha~=~.83$), Acceptance ($\alpha~=~.61$), Religion ($\alpha~=~.83$), and Self-blame ($\alpha_1~=~.75$, $\alpha_2~=~.71$). \par
\textbf{Loneliness} was measured using the $6$-item version of the De Jong Gierveld Loneliness Scale~\cite{gierveld2006}. The items are equally distributed among two factors, emotional; $\alpha_1~=~.68$, $\alpha_2~=~.69$) (\textit{e.g.}, ``I often feel rejected'') and social; $\alpha_1~=~.84$, $\alpha_2~=~.87$ (\textit{e.g.}, ``There are plenty of people I can rely on when I have problems''). Participants indicated how lonely they felt during the past week. Responses were given on a $5$-point scale ranging from $1$ (Not at all) to $5$ (Every day). \par
\textbf{Compliance} with official recommendations was measured using three items of a compliance scale~\cite{wolf2020importance}. The items are `Washing hands thoroughly with soap', `Staying at home (except for groceries and 1x exercise per day)' and `Keeping a 2m (6 feet) distance to others when outside.' Responses were given on a $7$-point scale ranging from $1$ (never complying to this guideline) to $7$ (always complying to this guideline, $\alpha~=~.71$). \par
 
\textbf{Anxiety} was measured using an adapted version of the $7$-item Generalized Anxiety Disorder scale~\cite{Spitzer2006GAD7}. Participants indicate how often they have experienced anxiety over the past week to different situations. Example questions are ``Feeling nervous, anxious, or on edge'' and ``Not being able to stop or control worrying''. Responses were given on a 5-point scale ranging from $1$ (Not at all) to $5$ (Every day, $\alpha_1~=~.93$, $\alpha_2~=~.93$). Additionally, we measured specific COVID-19 and future pandemic related concerns with two items ``How concerned do you feel about COVID-19?'' and ``How concerned to you about future pandemics?'' Responses on this were given by a $5$-point scale ranging from $1$ (Not at all concerned) to $5$ (Extremely concerned; $\alpha~=~.82$)~\cite{Nelson2020psychological}. \par
\textbf{Stress} was measured using a four-item version of the Perceived Stress Scale~\cite{Cohen1988perceived}. Participants indicate how often they experienced stressful situations in the past week. Example items include ``In the last week, how often have you felt that you were unable to control the important things in your life?'' and ``In the last week, how often have you felt confident about your ability to handle your personal problems?''. Responses were registered on a $4$-point scale ranging from $1$ (Never) to $4$ (Very often; $\alpha_1~=~.80$, $\alpha_2~=~.77$). \par
\textbf{Boredom} was measured using the $8$-item version~\cite{struk2017short} of the Boredom Proneness Scale~\cite{farmer1986boredom}. Example items include ``It is easy for me to concentrate on my activities'' and ``Many things I have to do are repetitive and monotonous''. Responses were on a $4$-point Likert scale ranging from $1$ (Strongly disagree) to $7$ (Strongly agree; $\alpha_1~=~.87$, $\alpha_2~=~.87$). \par
\textbf{Daily Routines} was measured with five items: ``I am planning a daily schedule and follow it'', ``I follow certain tasks regularly (such as meditating, going for walks, working in timeslots, etc.)'', ``I am getting up and going to bed roughly at the same time every day during the past week'', ``I am exercising roughly at the same time (\textit{e.g.}, going for a walk every day at noon)'', and ``I am eating roughly at the same time every day''. Responses were taken on a $7$-point Likert scale ranging from $1$ (Does not apply at all) to $7$ (Fully applies; $\alpha_1~=~.75$, $\alpha_2~=~.78$). \par
\textbf{Conspiracy beliefs} was measured with a $5$-item scale as designed by ourselves for this study. The first two items were adapted from the Flexible Inventory of Conspiracy Suspicions~\cite{wood2017conspiracy}, whereas the latter three are based on more specific conspiracy beliefs: ``The real truth about Coronavirus is being kept from the public.'', ``The facts about Coronavirus simply do not match what we have been told by `experts' and the mainstream media'', ``Coronavirus is a bio-weapon designed by the Chinese government because they are benefiting from the pandemic most'', ``Coronavirus is a bio-weapon designed by environmental activists because the environment is benefiting from the virus most'', and ``Coronavirus is just like a normal flu''. Responses were collected on a $7$-point Likert scale ranging from $1$ (Totally disagree) to $7$ (Totally agree, $\alpha~=~.83$). \par
\textbf{Extraversion} was measured using the $4$-item extraversion subscale of the Brief HEXACO Inventory~\cite{DeVries2013HEXACO}. Responses were given on a $5$-point Likert scale ranging from $1$ (Strongly disagree) to $5$ (Strongly agree; $\alpha_1~=~.71$, $\alpha_2~=~.69$). 
Low scores on extraversion are an indication of introversion.
Since we found at wave 1 that extraversion and well-being were positively correlated contrary to our hypothesis (see below), and, in our view, contrary to widespread expectations, we decided to measure in wave 2 what participants' views are regarding the association between extraversion and well-being. We measured expectations with one item: ``Who do you think struggles more with the current pandemic, introverts or extraverts?'' Response options were `Introverts', `Both around the same', and `extraverts'. \par
\textbf{Autonomy, competence, and relatedness} needs of the self-determination theory~\cite{ryan2000self} was measured using the $18$-item balanced measure of psychological needs scale~\cite{sheldon2012balanced}. Example items include ``I was free to do things my own way' (need for autonomy; $\alpha_1~=~.72$, $\alpha_2~=~.76$), ``I did well even at the hard things'' (competence; $\alpha_1~=~.77$, $\alpha_2~=~.77$), and ``I felt unappreciated by one or more important people'' (recoded; relatedness; $\alpha_1~=~.79$, $\alpha_2~=~.78$). Participants were asked to report how true each statement was for them in the past week. Responses were given on a $5$-point scale ranging from $1$ (no agreement) to $5$ (much agreement). \par
\textbf{Extrinsic and intrinsic work motivation} was measured with the $6$-item extrinsic regulation 3-item and intrinsic motivation subscales of the Multidimensional Work Motivation Scale~\cite{gagne2015multidimensional}. The extrinsic regulation subscale measures social and material regulations. Specifically, participants were asked to answer some questions about why they put effort into their current job. Example items include ``To get others' approval (\textit{e.g.}, supervisor, colleagues, family, clients ...)'' (social extrinsic regulation; $\alpha~=~.85$), ``Because others will reward me ﬁnancially only if I put enough effort in my job (\textit{e.g.}, employer, supervisor...)'' (material extrinsic regulation; $\alpha~=~.71$) and ``Because I have fun doing my job'' (intrinsic motivation; $\alpha~=~.94$). Responses were given on a $7$-point scale ranging from $1$ (not at all) to $7$ (completely). \par
\textbf{Mental exercise} was measured with two items: ``I did a lot to keep my brain active'' and ``I performed mental exercises (\textit{e.g.}, Sudokus, riddles, crosswords)''. Participants indicated the extent to which the items were true for them in the past week on a $7$-point scale ranging from $1$ (Not at all) to $7$ (Very; $\alpha~=~.56$). \par
\textbf{Technical skills} was measured with one item: ``How well do your technological skills equip you for working remotely from home?'' Responses were given on a $7$-point scale ranging from $1$ (Far too little) to $7$ (Perfectly).

\subsection{Physiological factors} 
\label{ssec:Physiological}

\textbf{Diet} was measured with two items~\cite{ESS2014round7}: ``How often do you eat fruit, excluding drinking juice?'' and ``How often do you eat vegetables or salad, excluding potatoes?''. Responses were given on a $7$-point scale ranging from $1$ (Never) to $7$ (Three times or more a day; $\alpha~=~.60$) \par
\textbf{Quality of sleep} was measured with one item: ``How has the quality of your sleep overall been in the past week?'' Responses were given on a $7$-point scale ranging from $1$ (very low) to $7$ (perfectly). \par
\textbf{Physical activity} was measured with an adapted version of the $3$-item Leisure Time Exercise Questionnaire~\cite{godin1985simple}. Participants were be asked to report how many hours in the past they have been mildly, moderately, and strenuously exercising. The overall score was computed as followed~\cite{godin1985simple}: $3 \times$ mild + $5 \times$ moderate + $9 \times$ strenuously. Missing responses for one or more of the exercise types were be treated as $0$. \par

\subsection{Social factors} 
\label{ssec:independent2}

\textbf{Quality and quantity of social contacts outside of work} were measured with three items. We adapted two items from the social relationship quality scale~\cite{birditt2007relationship} and added one item to measure the quantity: ``I feel that the people with whom I have been in contact over the past week support me'', ``I feel that the people with whom I have been in contact over the past week believe in me'', and ``I am happy with the amount of social contact I had in the past week.'' Responses were given on a $6$-point Likert scale ranging from $1$ (Strongly disagree) to $6$ (Strongly agree; $\alpha_1~=~.73$, $\alpha_2~=~.77$). \par 
\textbf{Volunteering} was measured with three items that measure people's behavior over the past week: ``I have been volunteering in my community (\textit{e.g.}, supported elderly or other people in high-risk groups)'', ``I have been supporting my family (\textit{e.g.}, homeschooling my children)'' and ``I have been supporting friends, and family members (\textit{e.g.}, listened to the worries of my friends)''. Responses were given on a 7-point scale ranging from $1$ (Not at all) to $7$ (Very often; $\alpha~=~.45$). \par
\textbf{Quality and quantity of communication with colleagues and line managers} was measured with three items: ``I feel that my colleagues and line manager have been supporting me over the past week'', ``I feel that my colleagues and line manager believed in me over the past week'', and ``Overall, I am happy with the interactions with my colleagues and line managers over the past week.'' Responses were given on a $6$-point Likert scale ranging from $1$ (Strongly disagree) to $6$ (Strongly agree; $\alpha_1~=~.88$, $\alpha_2~=~.92$). \par

\subsection{Situational factors and demographics} 
\label{ssec:independent3}

\textbf{Distractions at home} was measured with two items: ``I am often distracted from my work (\textit{e.g.}, noisy neighbors, children who need my attention)'' and ``I am able to focus on my work for longer time periods'' (recoded). Responses were given on a $5$-point scale ranging from $1$ (Not at all) to $5$ (Very often; $\alpha_1~=~.64$, $\alpha_2~=~.63$). \par 
Whether participants lived alone or with other \textbf{people} was assessed by asking them how many Babies, Toddlers, Children, Teenagers, and Adults participants were currently living with. We asked for the specific five groups separately because it allowed us to explore whether, for example, toddlers had a different impact on well-being and productivity than teenagers. However, the number of babies, toddlers, children, teenagers, and adults the participants were living with was uncorrelated to their well-being and productivity, $r$s $\leq .19$. Therefore, we summed them up into one variable, which we called people (\textit{i.e.}, the number of people the participant was living with). \par
\textbf{Financial security} was measured with two items that reflect the current but also the expected financial situation~\cite{glei2019growing}: ``Using a scale from 0 to 10 where 0 means `the worst possible financial situation' and 10 means `the best possible financial situation', how would you rate your financial situation these days?'' and ``Looking ahead six months into the future, what do you expect your financial situation will be like at that time?''. Responses were given on a 11-point scale ranging from $0$ (the worst possible financial situation) to $10$ (the best possible financial situation; $\alpha~=~.81$). \par
\textbf{Office set-up} was measured with three items: ``In my home office, I do have the technical equipment to do the work I need to do (\textit{e.g.}, appropriate PC, printer, stable and fast internet connection)'', ``On the computer or laptop I use while working from home I do have the software and access rights I need'', and `My office chair and desk are comfortable and designed to prevent back pain or other related issues''. Responses were given on a $7$-point Likert scale ranging from $1$ (Strongly disagree) to $7$ (Strongly agree; $\alpha~=~.65$). \par 
\textbf{Demographic information} were assessed with the following items: ``What is your gender?'', ``How old are you?'' ``What type of organization do you work in'' (public, private, unsure, other), ``What is your yearly gross income?'' (US\$$<$$20,000$, US\$$20-40,000$, US\$$40.001-60,000$, US\$$60,001-80,000$, US\$$80,001-100,000$, $>$US\$$100,000$; converted to the participant's local currency), ``In which country are you based?'', ``What percentage of your time have you been working remotely (\textit{i.e.}, not physically in your office) over the past 12 months?'', ``In which region/state and country are you living?'', ``Is there still a lockdown where you are living?''.

\section{Analysis}
\label{sec:analysis}
The data analysis consists of two parts. First, we used the data from time 1 to identify the variables that explain variance in participant well-being and productivity beyond the other variables. Second, we used the Pearson product-moment correlation coefficient ($r$) to identify which variables were correlated with at least $r$~=~$.30$ with well-being and productivity, to test whether they predict our two outcomes over time. 
$r$ is an effect size which expresses the strength of the linear relation between two variables.
We used $.30$ as a threshold as we are interested in identifying variables correlated with at least a medium-sized magnitude~\cite{cohen1992power} with one or both of our outcome variables. Also, a correlation of $\geq$ $.30$ indicates that the effect is among the top $25$\% in individual difference research~\cite{gignac2016effect}. Finally, selecting an effect size of this magnitude provides an effective type-I error control, as in total, we performed $103$ correlation tests at time 1 alone ($51$ independent variables correlated with the two dependent variables, which were also correlated among each other). Given a sample size of $192$, this effectively changes our alpha level to $.0001$, which is conservative. This means that it is improbable that we erroneously find an effect in our sample even though there is no effect in the population (\textit{i.e.}, commit the type-I or false-positive error) \par
We neither transformed the data for any analysis nor added any control variables\footnote{Adding control variables without a good justification can increase the type-I error rate \cite{simmons2011false}. However, we run additional analyses on the following demographic information: age, gender, and country. They were not associated with any of the two outcome variables and only correlated with one of the predictor variables (see Tables \ref{tab:correlations}, \ref{tab:womenmen}, \ref{tab:country}, and \ref{tab:cor1_complete}).} Unless otherwise indicated above, scales were formed by averaging the items. The collected dataset is publicly available to support other researchers in understanding the impact of (enforced) work-from-home policies. 

\subsection{Analysis of time 1 data}
To test which of the variables listed in Figure~\ref{fig:model} explains unique variance in well-being and productivity, we performed two multiple regression analyses with all variables that were correlated with the two outcome variables with $\geq .30$. In the first analysis, well-being is the dependent variable; in the second analysis, we use productivity as the dependent variable. This allows us to identify the variables that explain unique variance in the two dependent variables. However, one potential issue of including many partly correlated predictors is multicollinearity, which can lead to skewed results. If the Variance Inflation Factor (VIF) is larger than $10$, multicollinearity is an issue~\cite{Chatterjee1991regression}. Therefore, we tested whether the variance inflation factor would exceed $10$ before performing any multiple regression analysis.  

\subsection{Analysis of longitudinal data}
To analyze the data from both time-points, we performed a series of structural equation modeling analyses with one predictor variable and one outcome variable using the R-package \textit{lavaan}~\cite{rosseel2012lavaan}. Unlike many other types of analyses, structural equation modeling adjusts for reliability~\cite{westfall2016statistically}. Specifically, models were designed with one predictor (\textit{e.g.}, stress) and one outcome (\textit{e.g.}, well-being) both as measured at time 1 and at time 2. We allowed autocorrelations (\textit{e.g.}, between well-being at time 1 and at time 2) and cross-paths (\textit{e.g.}, between stress at time 1 and well-being at time 2). Autocorrelations are essential because, without them, we might erroneously conclude that, for example, stress at time 1 predicts well-being at time 2, although it is the part of stress which overlaps with well-being, which predicts well-being at time 2~\cite{rogosa1980critique}. To put it simply, we can only conclude that X1 predicts Y2 if we control for Y1. No items or errors were allowed to correlate. This is usually done to improve the model fit but has also been criticized as atheoretical: To determine which items and errors should be allowed to correlate to improve model fit can only be done after the initial model is computed. Therefore, it is a data-driven approach which emphasizes too much on the model fit~\cite{gana2019structural,hermida2015problem,maccallum1992model}. The regression (or path) coefficients and associated $p$-values were not affected by the estimator type. We compared in our analyses the standard maximum likelihood (ML), the robust maximum likelihood (MLR), and the multi-level (MLM) estimator. As fit indices, we report the CFI, RMSEA, and SRMR. To assess whether the fit indices are sufficient (i.e., from which point onward the data fits well to the model), we relied on the following cut-off values~\cite{hair2006multivariate,kline2015principles}: CFI $\geq$ .90, and RMSEA and SRMR $\leq$ .08.

\section{Results}
\label{sec:results}

\subsection{Correlations}
The pattern of correlations was overall consistent with the literature. At time 1, $16$ variables were correlated with well-being at $r \geq .30$ (Tables~\ref{tab:correlations} and~\ref{tab:cor1_complete})\footnote{The Pearson's correlation coefficient ($r$) represents the strength of a linear association between two variables and can range between $-1$ (perfect negative linear association), $0$ (no linear association), to $1$ (perfect positive linear association). The regression coefficient B indicates how much the outcome changes if the predictor increases by one unit. For example, the B of stress (the dependent variable or predictor)} predicting well-being (the independent variable or outcome) is $-.60$. This indicates that a person who has a well-being level of $6$ has a stress level that is of $-.60$ units lower than a person who has a well-being level of $5.$. Stress, $r~=~-.58,$ quality of social contacts, $r~=~.49$, and need for autonomy, $r~=~.48$ were strongest associated with well-being (all $p$ $<$ .0001). The pattern of results from the $14$ coping strategies was also in line with the literature~\cite{carver1989assessing}: self-blame, $r~=~-.36, p~<~.001$, behavioral disengagement, $r~=~-.31, p~<~.001$, and venting $r~=~-.28, p~<~.001$ were negatively correlated with well-being.  Interestingly, generalized anxiety was more strongly associated with well-being than COVID-19 related anxiety ($r~=~-.46$ vs. $-.25$), which might suggest that specific worries have a less negative impact on well-being\footnote{A multiple regression with generalized anxiety and COVID-19 related anxiety supports this interpretation: Only generalized anxiety, $B~=~-.58, SE~=~.10, p~<~.001$, but not COVID-19 related anxiety, $B~=~-.11, SE~=~.09, p~=~.21$. This suggests that whether people are worried about COVID-19 specifically has little impact on their well-being. Their general level of anxiety matters substantially.}. This also suggests that our findings are at least partly COVID-19 independent; namely, if people were terrified by this virus, COVID-19 related anxiety would have been a stronger predictor than generalized anxiety.

Contrary to our expectations, extraversion was positively correlated with well-being, both at waves 1 and 2. The pattern of the associations was similar at time 2. A reason for participants' misinterpretation of the intensity to struggle with working from home for introverts could be explained by introverts usually having to avoid unwanted social interactions, and due to being quarantined, they now have to put effort into having social interactions actively. The added challenge to contribute more energy than usual to not being too lonely and changing their usual behavioral pattern demands much more from introverts than extraverts \cite{davidson2015introversion,wei2020social}. 
\\
At time 1, four variables were correlated with productivity at $r \geq .30$ (Tables~\ref{tab:correlations} and~\ref{tab:cor1_complete}: Need for competence, $r~=~-.37$, distractions, $r~=~-.34$, boredom, $r~=~-.33$, and communication with colleagues and line-managers $r~=~.30$. Surprisingly, work motivations were uncorrelated with well-being at $\alpha~=~.001$. At time 2, only distraction was still correlated with productivity, $r~=~-.26, p~<~.001$ (see also Table~\ref{tab:cor2_complete}). The strength of association of most variables with productivity dropped between time 1 and 2, which means that those variables associated with productivity at wave 1 were no longer or less strongly associated with productivity at wave 2. The strengths of correlations remained the same when we computed Spearman's rank correlation coefficients rather than Pearson's correlations (Spearman's coefficient is a non-parametric version of Pearson's $r$ and ranges also between $-1$ and $1$, see Tables \ref{tab:cor1_complete} and \ref{tab:cor2_complete}).

\subsubsection{Additional analysis regarding extraversion}

The counter-intuitive finding that well-being and extraversion are positively correlated surprised us. 
Thus, we added additional questions at time 2 to better understand this phenomenon.
The purpose of this further investigation is only to provide a more nuanced interpretation of the results of our quantitative analysis; it is not a stand-alone research about extraversion during the lockdown.

Interestingly, the finding that extraversion is positively correlated with well-being during lockdown is contrary to most participants' expectations. When asked whether introverts or extraverts struggle more with the COVID-19 pandemic, only $2$ participants correctly predicted introverts, where $136$ stated extraverts, with $46$ participants believing that both groups struggle equally. This highlights the value of our research because people's intuition can be blatantly wrong.

The explanation became more articulated through an analysis of the participants' statements about the informant's (I) choice. 
We now report selected quotes from participants, including their level of extraversion, in wave 1\footnote{Scores close to $1$ are indicative of an introverted personality trait, while $5$ of an extraverted one.}.
Some informants reported their direct experience supporting the feeling that extraverts struggle more than introverts.

\begin{displayquote}
``\textit{I'm introverted, and I don't feel the pandemic has affected me at all. Rules aren't hard to follow and haven't feel bad. I feel for extraverts; they would struggle a bit with the rules.}'' [I-101, extraversion score=$2.75$]
\end{displayquote}

\begin{displayquote}
``\textit{I'm an extravert; my wife is an introvert. I'm really struggling. She's fine.}'' [I-92, extraversion score = $5.00$]
\end{displayquote}

Nonetheless, a minority of participants also provide alternative interpretations.
According to those, both introverts and extraverts have difficulties in reaching out to people, although in different ways.
The motivation for such answers is that both personality types struggle with different challenges.

\begin{displayquote}
``\textit{Both types need company, just that each needs company on their own terms. Introverts prefer deeper contact with fewer people and extraverts less deep contact with a greater number of people.}'' [I-80, extraversion score = $3.75$]
\end{displayquote}

\begin{displayquote}
``\textit{Extraverts miss human contact; introverts find it even harder to mark their presence online (\textit{e.g.}, in meetings).}'' [I-160, extraversion score = $3.50$]
\end{displayquote}

Interestingly, there is one informant which provide an insightful interpretation, aligned with our results.

\begin{displayquote}
``\textit{Introverts usually have more difficulty communicating with others, and confinement worsens the situation because they will not try to talk to others through video conferences.}'' [I-136, extraversion score = $2.75$]
\end{displayquote}

The lack of a structured working setting, where introvert are routinely involved, causes further isolation.
Being `forced' to work remotely significantly increased difficulty in engaging with social contacts.
This means that introverts have to put much more effort into interacting with others instead of their typical behavior of reduced interaction in office-based environments.
Whereas extraverts have it easier to find some way to maintain their social contacts, introverts might struggle more.
Thus, the lockdown had a more negative impact on the well-being of introverts than of extraverts, as shown in Table~\ref{tab:correlations}.

\protected\def\stars#1{$^{#1}$}

\begin{table}[]
\sisetup{
    group-digits=true,
    detect-weight=true,
    detect-shape=true,
    table-format=-1.2,
    table-alignment = left,
    table-align-text-post=false,
    table-space-text-post = \stars{***}
}
\caption{Correlations $r$ at time 1 and 2, unstandardized regression coefficients $B$, and test-retest reliabilities $r\textsubscript{it}$}
\label{tab:correlations}
\resizebox{\textwidth}{!}{%
\begin{tabular}{@{}lSSSSSSSSS[table-format = 1.2]@{}}
\toprule
                           & \textbf{r\textsubscript{WB1}} & \textbf{B\textsubscript{WB1} } & \textbf{r\textsubscript{PR1}} & \textbf{B\textsubscript{PR1} } & \textbf{r\textsubscript{WB2}} & \textbf{B\textsubscript{WB2}} & \textbf{r\textsubscript{PR2}} & \textbf{B\textsubscript{PR2}} & \textbf{r\textsubscript{it}} \\ \midrule
Well-being (WB)            & 1.00          &               & .18**         &               & 1.00        &               & .20**       &               & .72***     \\
Productivity (PR)          & .18*          &               & 1.00          &               & .20**       &               & 1.00        &               & .50***     \\
Boredom                    & -.42***       & -.05         & -.33***       & -.05        & -.33***     & .14         & -.15*       & -.02        & .69***     \\
Behavioral-disengagement   & -.31***       & .12          & -.15*         &               & -.41***     & -.03        & -.08        &               & .54***     \\
Self-blame                 & -.36***       & .01          & -.21**        &               & -.40***     & -.08        & -.07        &               & .61***     \\
Relatedness                & .47***        & .03          & .22**         &               & .48***      & -.04        & .05         &               & .71***     \\
Competence                 & .41***        & -.20         & .37***        & .09         & .38***      & -.33*       & .22**       & .07         & .65***     \\
Autonomy                   & .48***        & .20          & .17*          &               & .54***      & .35*        & .09         &               & .76***     \\
Communication              & .41***        & .07          & .30***        & .04         & .39***      & .03         & .19**       & .02         & .67***     \\
Stress                     & -.58***       & -.60***      & -.27***       &               & -.54***     & -.34*       & -.08        &               & .73***     \\
Daily-routines             & .37***        & .12*         & .25***        &               & .42***      & .05         & .11         &               & .73***     \\
Distractions               & -.23**        & .06          & -.34***       & -.06        & -.33***     & .00         & -.26***     & -.08        & .63***     \\
Generalized-anxiety        & -.46***       & .01          & -.21**        &               & -.53***     & -.07        & -.09        &               & .76***     \\
Emotional-loneliness       & -.41***       & -.13         & -.23**        &               & -.45***     & -.14        & -.16*       &               & .72***     \\
Social-loneliness          & -.37***       & .08          & -.13          &               & -.47***     & -.01        & -.08        &               & .69***     \\
Quality of social contacts & .49***        & .22*         & .24***        &               & .53***      & .30**       & .12         &               & .66***     \\
Extraversion               & .32***        & .22          & .24***        &               & .28***      & -.00        & .08         &               & .74***     \\
Quality-of-Sleep           & .42***        & .05          & .27***        &               & .48***      & .14*        & .14         &               & .76***     \\
Conspiracy                 & -.04          &               & .01           &               &               &               &               &               &              \\
Self-distraction           & -.12          &               & .06           &               &               &               &               &               &              \\
Active-coping              & .22**         &               & .05           &               &               &               &               &               &              \\
Denial                     & -.12          &               & .00           &               &               &               &               &               &              \\
Substance-use              & -.08          &               & -.11          &               &               &               &               &               &              \\
Emotional-support          & .10           &               & -.04          &               &               &               &               &               &              \\
Instrumental-support       & -.09          &               & -.11          &               &               &               &               &               &              \\
Venting                    & -.28***       &               & -.15*         &               &               &               &               &               &              \\
Positive-reframing         & .19**         &               & -.06          &               &               &               &               &               &              \\
Planning                   & -.09          &               & -.09          &               &               &               &               &               &              \\
Humor                      & .07           &               & -.13          &               &               &               &               &               &              \\
Acceptance                 & .20**         &               & .01           &               &               &               &               &               &              \\
Religion                   & -.12          &               & -.18*         &               &               &               &               &               &              \\
Office-setup               & .14           &               & .10           &               &               &               &               &               &              \\
Self-Control               & .26***        &               & .17*          &               &               &               &               &               &              \\
Volunteering               & .07           &               & .01           &               &               &               &               &               &              \\
Diet                       & .17*          &               & .16*          &               &               &               &               &               &              \\
Exercising-overall         & .10           &               & .00           &               &               &               &               &               &              \\
Financial-situation        & .27***        &               & .19**         &               &               &               &               &               &              \\
Covid19-anxiety            & -.25***       &               & .13           &               &               &               &               &               &              \\
Mental-exercise            & .25***        &               & .18*          &               &               &               &               &               &              \\
Extrinsic-social           & -.10          &               & -.04          &               &               &               &               &               &              \\
Extrinsic-material         & -.22**        &               & -.13          &               &               &               &               &               &              \\
Intrinsic-motivation       & .26***        &               & .22**         &               &               &               &               &               &              \\
People                     & .03           &               & .09           &               &               &               &               &               &              \\
Compliance                 & .05           &               & .13           &               &               &               &               &               &              \\
Technological-Skills       & .24***        &               & .19**         &               &               &               &               &               &              \\
Time-remote                & -.06          &               & -.04          &               &               &               &               &               &              \\
Age                        & -.06          &               & .07           &               &               &               &               &               &              \\ \bottomrule
\end{tabular}%
}
\small \textit{Note}. $r$: correlation, $B$: unstandardized regression estimate, r\textsubscript{it}: test-retest correlation,  WB1: Well-being at time 1 (\textit{e.g.}, correlations for well-being with other variables), PR2: Productivity at time 2 (\textit{e.g.}, unstandardized regression estimate of the four variables predicting productivity in a linear multiple regression). 95\%-confidence intervals, 99.9\% confidence intervals, and Spearman's rho correlation are displayed in Tables \ref{tab:cor1_complete} and \ref{tab:cor2_complete}. \\
{\scriptsize{Signif. codes: $^{***}<.001$, $^{**}<0.01$, $^{*}<0.05$, $.<0.1$}}
\end{table}

\subsection{Unique influence  ---  Multiple regression analyses}
To test which of the predictors had a unique influence on well-being and productivity, we included all variables that were correlated with either outcome with at least .30 at time 1. This is a conservative test because many predictors are correlated among each other and thus taking variance from each other. Also, it allowed us to repeat the same analysis at time 2 because all predictors which correlated with either well-being or productivity at time 1 with $r \geq .30$ were included at time 2. In a first step, we tested whether multicollinearity was an issue. This was not the case, with VIF $<4.1$ for all four regression models and thus clearly below the often-used threshold of $10$~\cite{Chatterjee1991regression}. \par
Sixteen variables correlated with well-being $r \geq .30$ (Table~\ref{tab:correlations}). Together, they explained a substantial amount of variance in well-being at time 1, $R^2~=~.44, adj. R^2~=~.39, F(16, 167)~=~8.21, p~<~.0001$, and at time 2, $R^2~=~.47, adj. R^2~=~.42, F(16, 162)~=~8.90, p~<~.0001$. At time 1, stress (negatively), social contacts, and daily routines uniquely predicted well-being at $\alpha~=~.05$ (see Table~\ref{tab:correlations}, column 3, and Table~\ref{tab:wellbeing-one}). At time 2, need for competence and autonomy, stress, quality of social contacts, and quality of sleep uniquely predicted well-being at $\alpha~=~.05$ (see Table~\ref{tab:correlations}, column 7, and Table~\ref{tab:wellbeing-two}). Together, stress and quality of social contacts predicted at both time points significantly well-being. 
Four variables correlated with productivity $r \geq .30$ (Table~\ref{tab:correlations}). Together, they explained 16\% of variance in productivity at time 1, $R^2~=~.18, adj. R^2~=~.16, F(4, 179)~=~9.60, p~<~.0001$, and 8\% at time 2, $R^2~=~.08, adj. R^2~=~.06, F(4, 173)~=~4.02, p~=~.004$. At both time points, none of the four variables explained variance in productivity beyond the other three variables, suggesting that they all are associated with productivity but we lack statistical power to disentangle the effects (Tables~\ref{tab:productivity-one} and \ref{tab:productivity-two}). We also visualized the regression coefficients alongside their respective confidence intervals (see Figures~\ref{fig:reg_well_being1}, \ref{fig:reg_well_being2}, \ref{fig:reg_productivity1}, and \ref{fig:reg_productivity2} in the Appendix)

There is an ostensible discrepancy between some correlations and the estimates of the regression analyses which requires further explanations. An especially large discrepancy appeared for the variable need for competence, which correlated positively with well-being at time 1 and 2, $r$~=~$.41$ with $p$~$<$~$.001$, and $r$~=~$.38$ with $p$~$<$~$.001$, but was \textit{negatively} associated with well-being when controlling for other variables in both regression analyses, $B$~=~$-.20$, $p$~=~$.24$, and $B$~=~$-.33$, $p$~=~$.04$. This suggests that including a range of other variables, which serve as control variables, impact the results. Indeed, exploratory analyses revealed that need for competence was no longer associated with well-being when we included need for autonomy. That is, when we performed a multiple regression with the needs for autonomy and competence as the only predictors, need for competence became non-significant. Need for competence also includes an autonomy competent, which might explain this. It is easier to fulfill one's need for competence while being at least somewhat autonomous~\cite{ryan2000self}. Further, including generalized anxiety and boredom reversed the sign of the association: Need for competence became negatively associated with well-being. Including those two variables remove the variance that is associated with enthusiasm (boredom reversed) and courage (generalized anxiety reversed), which might explain the shift to negative association with well-being. Together, controlling for need for autonomy, generalized anxiety, and boredom, takes away positive aspects of need for competence, leaving a potentially cold side that might be closely related to materialism, which is negatively associated with well-being~\cite{dittmar2014relationship}.

\begin{table}[h]
\sisetup{
    group-digits=true,
    detect-weight=true,
    detect-shape=true,
    table-format=-1.3,
    table-alignment = left
}
\caption{Predictors of well-being wave 1}
\begin{tabular}{@{}p{4cm}SS[table-format=1.3]SS[table-align-text-post=false, table-space-text-post = \stars{***}]@{}}
  \toprule
 & \textbf{Estimate} & \textbf{Std. Error} & \textbf{t value} & \textbf{Pr($>|t|$)} \\ 
  \midrule
  Boredom                   & -0.047    & 0.100     & -0.474    & 0.636 \\ 
  Behavioral disengagement  & 0.120     & 0.112     & 1.073     & 0.285 \\ 
  Self blame                & 0.013     & 0.113     & 0.116     & 0.908 \\ 
  Relatedness               & 0.025     & 0.173     & 0.147     & 0.884 \\ 
  Competence                & -0.201    & 0.169     & -1.186    & 0.237 \\ 
  Autonomy                  & 0.203     & 0.171     & 1.188     & 0.237 \\ 
  Communication             & 0.073     & 0.106     & 0.690     & 0.491 \\ 
  Stress                    & -0.605    & 0.178     & -3.393    & 0.001*** \\ 
  Daily routines            & 0.125     & 0.061     & 2.038     & 0.043* \\ 
  Distractions              & 0.061     & 0.105     & 0.580     & 0.563 \\ 
  Generalized anxiety       & 0.010     & 0.146     & 0.071     & 0.944 \\ 
  Emotional loneliness      & -0.126    & 0.133     & -0.948    & 0.344 \\ 
  Social loneliness         & 0.082     & 0.108     & 0.761     & 0.447 \\ 
  Social contacts           & 0.224     & 0.106     & 2.125     & 0.035* \\ 
  Extraversion              & 0.223     & 0.127     & 1.757     & 0.081{.} \\ 
  Quality of Sleep          & 0.053     & 0.058     & 0.918     & 0.360 \\ 
   \bottomrule
   \multicolumn{2}{l}{\scriptsize{Signif. codes: $^{***}<.001$, $^{**}<0.01$, $^{*}<0.05$, $.<0.1$}}
\end{tabular}
\label{tab:wellbeing-one}

\vspace{1cm}

\caption{Predictors of productivity wave 1}
\sisetup{
    group-digits=true,
    detect-weight=true,
    detect-shape=true,
    table-format=-1.3,
    table-alignment = left
}
\begin{tabular}{@{}p{4cm}SS[table-format=1.3]SS[table-align-text-post=false, table-space-text-post = \stars{***}]@{}}
  \toprule
 & \textbf{Estimate} & \textbf{Std. Error} & \textbf{t value} & \textbf{Pr($>|t|$)} \\ 
  \midrule
  Boredom &        -0.053 & 0.031 & -1.675   & 0.096{.} \\ 
 Competence &     0.088  & 0.053 & 1.650    & 0.101 \\ 
 Communication &  0.043  & 0.034 & 1.256    & 0.211 \\ 
 Distractions &   -0.065 & 0.036 & -1.795   & 0.074{.} \\ 
 \bottomrule
   \multicolumn{2}{l}{\scriptsize{Signif. codes: $^{***}<.001$, $^{**}<0.01$, $^{*}<0.05$, $.<0.1$}}
\end{tabular}
\label{tab:productivity-one}
\end{table}

\begin{table}[h]
\sisetup{
    group-digits=true,
    detect-weight=true,
    detect-shape=true,
    table-format=-1.3,
    table-alignment = left
}
\caption{Predictors of well-being wave 2}
\begin{tabular}{@{}p{4cm}SS[table-format=1.3]SS[table-align-text-post=false, table-space-text-post = \stars{***}]@{}}
 \toprule
 & \textbf{Estimate} & \textbf{Std. Error} & \textbf{t value} & \textbf{Pr(($>|t|$)} \\ 
  \midrule
Boredom                      & 0.144    & 0.094     & 1.529     & 0.128 \\ 
Behavioral disengagement     & -0.035   & 0.140     & -0.249    & 0.804 \\ 
Self blame                   & -0.075   & 0.145     & -0.518    & 0.605 \\ 
Relatedness                  & -0.036   & 0.156     & -0.228    & 0.820 \\ 
Competence                   & -0.329   & 0.159     & -2.068    & 0.040* \\ 
Autonomy                     & 0.347    & 0.146     & 2.380     & 0.018* \\ 
Communication                & 0.033    & 0.087     & 0.382     & 0.703 \\ 
Stress                       & -0.337   & 0.157     & -2.153    & 0.033* \\ 
Daily routines               & 0.046    & 0.064     & 0.728     & 0.467 \\ 
Distractions                 & 0.005    & 0.108     & 0.046     & 0.963 \\ 
Generalized anxiety          & -0.073   & 0.134     & -0.549    & 0.583 \\ 
Emotional loneliness         & -0.136   & 0.126     & -1.076    & 0.283 \\ 
Social loneliness            & -0.011   & 0.126     & -0.085    & 0.932 \\ 
Social contacts              & 0.304    & 0.114     & 2.676     & 0.008** \\ 
Extraversion                 & -0.001   & 0.114     & -0.011    & 0.991 \\ 
Quality of Sleep             & 0.144    & 0.056     & 2.576     & 0.011*\\ 
\bottomrule
   \multicolumn{2}{l}{\scriptsize{Signif. codes: $^{***}<.001$, $^{**}<0.01$, $^{*}<0.05$, $.<0.1$}}
\end{tabular}
\label{tab:wellbeing-two}

\vspace{1cm}

\caption{Predictors of productivity wave 2}
\sisetup{
    group-digits=true,
    detect-weight=true,
    detect-shape=true,
    table-format=-1.3,
    table-alignment = left
}
\begin{tabular}{@{}p{4cm}SS[table-format=1.3]SS[table-align-text-post=false, table-space-text-post = \stars{***}]@{}}
 \toprule
 & \textbf{Estimate} & \textbf{Std. Error} & \textbf{t value} & \textbf{Pr(($>|t|$)} \\ 
  \midrule
  Boredom         & -0.015 & 0.032 & -0.479   & 0.632 \\ 
Competence      & 0.065  & 0.060 & 1.089    & 0.278 \\ 
Communication   & 0.021  & 0.032 & 0.662    & 0.509 \\ 
Distractions    & -0.077 & 0.041 & -1.874   & 0.063{.} \\ 
\bottomrule
   \multicolumn{2}{l}{\scriptsize{Signif. codes: $^{***}<.001$, $^{**}<0.01$, $^{*}<0.05$, $.<0.1$}}
\end{tabular}
\label{tab:productivity-two}
\end{table}

\subsection{Longitudinal analysis}
After identifying the independent variables that are more strongly related to well-being and productivity, we are now performing our longitudinal analysis, which will allow us to assess whether any of our sixteen predictors or independent variables predict one of our dependent variables at time 2 or is predicted by it.
Test-retest reliabilities were satisfactory for all variables, supporting our data's quality (last column of Table~\ref{tab:correlations}).

\subsubsection{Structural Equation Modeling}

In total, we performed $20$ structural equation modeling (SEM) analyses to test whether well-being and productivity are predicted by or predict any of the 16 independent variables for well-being, including one model in which we tested whether well-being predicts productivity or vice versa, and four models for productivity. Since the probability of a false positive is very high, due to the high number of models analyzed, we used a conservative error rate of $.005$. We are using a different threshold for the longitudinal analysis than for the correlation analyses since we did a different number of tests for the latter. Occasionally, the model fit indices indicated that the data did not fit well to the models (cf. Table~\ref{tab:sem}, last three columns). This was especially the case for the models, including the need for autonomy, competence, and relatedness, which we do not discuss further.\footnote{As noted above, we did not use the modification indices to increase the model fit (\textit{e.g.}, allow items to correlate). Nevertheless, we explored the impact of improving the model fit based on the modification indices for the model, including well-being and social loneliness. We chose this model because it contained the smallest $p-value$ (B = .124, p = .006, which was just above our .005 threshold). Allowing three well-being items to covary between t1 and t2 increased unsurprisingly the model fit substantially (\textit{e.g.}, CFI = .969 $\rightarrow$ .992, RMSEA = .057 $\rightarrow$ .029). However, the coefficient $B$ decreased slightly to .120, and the $p-value$ increased to .009. Together, this suggests that improving the model fit does not impact the regression coefficients, and therefore not impact our conclusion not to reject the null-hypothesis.}

\begin{figure}[h]
\centering
\includegraphics[width=1\linewidth]{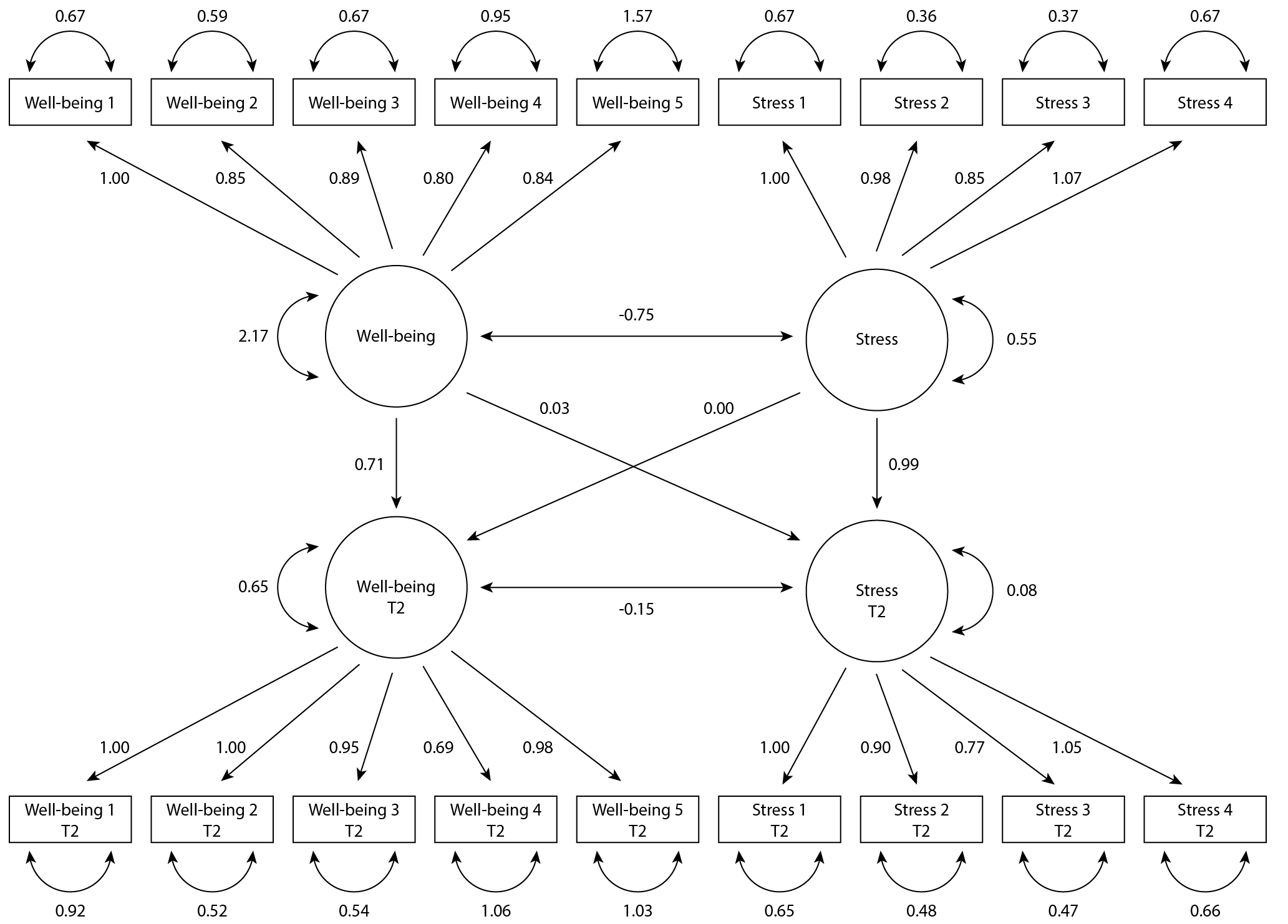}
\caption{SEM analysis of stress and well-being in wave 1 and 2}
\label{fig:sem2}
\end{figure}

One example of our SEM analyses is presented in Figure~\ref{fig:sem2}, where we looked at the causal relations between stress and well-being in waves 1 and 2. The boxes represent the items and the circles the variables (\textit{e.g.}, stress). The arrows between the items and the variables represent the loadings, that is, how strongly each of the items contributes to the overall variable score (\textit{e.g.}, item 3 of the stress scale contributes least and item 4 most to the overall score at both time points). The circular arrows represent errors. The bidirectional arrows between the variables represent the covariances, which are comparable to correlations. The one-handed arrows show causal impacts over time. The arrows between the same variables (\textit{e.g.}, well-being 1, and well-being 2) show how strongly they impact each other and are comparable to the test-retest correlations. The most critical arrows are those between well-being 1 and stress 2 and between stress 1 and well-being 2. They show whether one variable causally predicts the other.

To provide a better understanding of our SEM analyses, we will guide the reader through the example shown in Figure~\ref{fig:sem2}.
The values (of this and all SEM analyses) are displayed in Table~\ref{tab:sem}. Columns 2-4 of Table~\ref{tab:sem} show that stress and well-being were significantly associated at time 1, $B$~=~$-0.75$, $SE$~=~$.13$, $p~<~.001$. 
This association was mirrored at time 2, $B$~=~$-0.15$, $SE$~=~$.05$, $p$~=~.001 (columns 5-7). 
Columns 8-10 show that stress at time 1 did not significantly predict well-being at time 2, $B$~=~$-0.00$, $SE$~=~$.16$, $p$~=~$.99$. Columns 8-10 of the second part of Table~\ref{tab:sem} also show that well-being at time 1 did not predict stress at time 2, $B$~=~$0.03$, $SE$~=~$.05$, $p$~=~$.55$. 
Columns 2-4 of the second part show the autocorrelation of well-being, that is how strongly well-being at time 1 predicts well-being at time 2, $B$~=~$0.71$, $SE$~=~$.09$, $p~<~.001$. 
Autocorrelations can be broadly understood as the unstandardized version of the test-retest correlations (reliability) reported in Table~\ref{tab:correlations}. 
Columns 5-7 of the second part show the autocorrelation of stress, which are also significant $B$~=~$.99$, $SE$~=~$.16$, $p~<~.001$.
The last three columns indicate that the data fit reasonably well to the proposed model, $CFI = .93, RMSEA = .07, SRMR = .07$. It is worth noting that at time 1 (t1), the coefficient between well-being --- stress is $-.75$, at time 2 (t2) only $-.15$. 
This is likely an SEM artifact because the variances of both well-being and stress are larger at t1 than at t2: $2.17$ and $0.55$ at t1 vs. $0.65$ and $0.08$ at t2 (see the double-headed arrows in Figure~\ref{fig:sem2}). 
Because the standard errors also differ for the two coefficients, $.13$ at t1 vs $.05$ at t2 (cf. Table~\ref{tab:sem}), both coefficients are significant, $p \leq .001$. The correlation analysis supports this view, since well-being and stress are correlated with $r = -.58$ at t1, and with $r = -.54$ at t2 (cf. Table~\ref{tab:correlations}), suggesting clearly that the relations between the two variables are very similar across both time points.

We conclude our SEM analyses by acknowledging that no model revealed any significant associations at $\alpha~=~.005$. Thus, no variable at time 1 (\textit{e.g.}, stress) is able to explain a significant amount of variance in another variable (\textit{e.g.}, well-being) at time 2.  
We only found a negative tendency regarding $Distraction$ $\rightarrow$ $Productivity$ with $B$~=~$-.154$, $p~=~.006$.

\subsubsection{Mixed Effects Modeling}

Additionally, we explored whether there are any mean changes between time 1 and 2, separately for all $18$ variables using mixed effects modeling. For example, has the well-being increased over time? This would suggest that people adapted further within a relatively short period of two weeks to the threat from COVID-19. Table~\ref{tab:within} shows that the arithmetic mean ($M$) of well-being has indeed slightly increased between time 1 and 2, $M$~=~4.14 vs. $M$~=~$4.34$. A closer look revealed that $91$ participants reported higher well-being at time 2 compared to time 1, $23$ reported the same level of well-being, and $70$ a lower level of well-being. Further, on average, people's score of behavioral disengagement and quality of social contacts increased, whereas emotional loneliness and the quality of communication with line managers and coworkers decreased. 

\subsection{Exploratory between Gender and Country Analyses}

Further, we tested for gender mean differences by comparing women and men across 65 variables (cf. Tab.~\ref{tab:womenmen}). Because of the large number of comparisons, we set our significance threshold to .001. With this threshold, only the coping strategy self-distraction resulted in significant differences with women reporting higher levels of it (\textit{e.g.}, "I've been turning to work or other activities to take my mind off things"). Other comparisons were in the expected direction but not statistically significant. For example, women tended to score higher on anxiety on average, which is in line with the literature~\cite{feingold1994gender}.

Finally, we explored whether there would be any mean differences between participants based in the United Kingdom ($n$ = 63) and the United States of America ($n$ = 52). We only selected those two countries because there were only 19 or fewer participants in each of the other countries. We again used a threshold of .001. With this threshold, only the work motivation material-extrinsic resulted in significant differences with people based in the USA reporting higher levels of it on average. This means that Americans are more driven by materialistic motivation (\textit{e.g.}, promotions, money) compared to UK professionals.

\begin{table}[]
\sisetup{
    group-digits=true,
    detect-weight=true,
    detect-shape=true,
    table-format=-1.3,
    table-alignment = left
}
\caption{Within-subject comparisons to analyze mean changes over time}
\resizebox{\textwidth}{!}{\begin{tabular}{@{}lllllSSSlll@{}}
\toprule
                             & \multicolumn{2}{l}{\textbf{Time 1}} & \multicolumn{2}{l}{\textbf{Time 2}} &            &                 &           &        &         &       \\ 
                             & M                & SD               & M                & SD               & \textit{t} & \textit{p}      & {Cohen’s d} & Higher & Smaller & Equal \\ \midrule
Well-being                   & 4.140             & 1.367            & 4.340             & 1.289            & -2.329     & 0.021           & -0.129    & 91     & 70      & 23    \\
Productivity                 & 0.990             & 0.419            & 1.032            & 0.436            & -1.575     & 0.117           & -0.116    & 87     & 77      & 19    \\
Boredom                      & 2.936            & 1.136            & 2.927            & 1.158            & -0.330      & 0.742           & -0.019    & 91     & 79      & 14    \\
Behavioral-disengagement     & 1.805            & 0.936            & 2.062            & 1.030             & -3.621     & {\textless} 0.001 & -0.256    & 82     & 40      & 62    \\
Self-blame                   & 1.812            & 0.990             & 1.880             & 1.013            & -0.962     & 0.337           & -0.062    & 60     & 52      & 72    \\
Need   for Relatedness       & 3.497            & 0.830             & 3.559            & 0.803            & -1.130      & 0.260           & -0.063    & 86     & 73      & 25    \\
Need   for Competence        & 3.572            & 0.735            & 3.582            & 0.731            & -0.040      & 0.968           & -0.002    & 82     & 82      & 20    \\
Need   for Autonomy          & 3.483            & 0.688            & 3.511            & 0.732            & -0.572     & 0.568           & -0.029    & 88     & 67      & 29    \\
Communication                & 4.534            & 0.996            & 4.292            & 1.185            & 3.244      & 0.001           & 0.199     & 57     & 81      & 38    \\
Stress                       & 2.501            & 0.807            & 2.520             & 0.797            & -0.593     & 0.554           & -0.032    & 81     & 64      & 39    \\
Daily   routines             & 4.681            & 1.561            & 4.717            & 1.533            & -0.108     & 0.914           & -0.006    & 71     & 72      & 41    \\
Distractions                 & 2.466            & 0.934            & 2.443            & 0.895            & 0.188      & 0.851           & 0.012     & 58     & 64      & 62    \\
Generalized   anxiety        & 2.245            & 1                & 2.174            & 1.010             & 1.246      & 0.214           & 0.064     & 69     & 90      & 25    \\
Emotional   loneliness       & 2.111            & 0.903            & 2.007            & 0.871            & 2.077      & 0.039           & 0.114     & 54     & 79      & 51    \\
Social   loneliness          & 2.641            & 1.004            & 2.563            & 1.017            & 0.807      & 0.421           & 0.047     & 65     & 79      & 40    \\
Quality   of social contacts & 4.109            & 1.093            & 4.312            & 1.077            & -2.612     & 0.010           & -0.159    & 91     & 54      & 39    \\
Extraversion                 & 3.448            & 0.786            & 3.457            & 0.778            & -0.195     & 0.846           & -0.009    & 73     & 59      & 52    \\
Quality of Sleep           & 4.130             & 1.754            & 4.174            & 1.686            & 0.310       & 0.757           & 0.016     & 54     & 51      & 79    \\ \bottomrule
\end{tabular}}
\scriptsize{Note. \textit{t}: \textit{t}-value of a paired sample $t$-test; Higher: Absolute number of people who scored higher on a variable at time 2 compared to time 1; Smaller}: Number of people who scored lower at time 2; Equal: People whose score has not changed over time.
\label{tab:within}
\end{table}

\vspace{1em}
\subsection{Conceptual replication analysis}

Our finding that office-setup is not significantly related to well-being and productivity seems to contradict a recent cross-sectional study by Ralph et al.~\cite{Ralph2020pandemic} that investigated how the fear of bioevents, disaster preparedness, and home office ergonomics predict well-being and productivity among software developers. In that study, ergonomics was positively related to both well-being and productivity. To measure ergonomics, the authors created six items concerning distractions, noise, lighting, temperature, chair comfort, and overall ergonomics. 
The first two items are closely related to our measure of distraction, which was negatively associated with well-being in wave 1 of our sample, $r$~=~-$.23$, and productivity, $r$= $-.34$. In contrast, the following four items are more closely associated with office-setup in our survey, which was positive but not significantly associated with well-being, $r$~=~$.14$, and productivity, $r$~=~$.10$. 

To better understand such inconsistency with our result, we run a replication analysis using Ralph et al.'s data.
To test whether ergonomics' effect is mainly driven by distraction and noise, we combined the first two items into variable ergonomics-distractions (recoded, higher scores indicate less distraction) and the other four items into ergonomics-others. Indeed, ergonomics distractions was more strongly correlated with well-being, $r$~=~$.25$, and productivity, $r$~=~$.29$, than was ergonomics-other, $r$s~=~$.19$ and $.19$, respectively. This suggests that our findings replicate those of Ralph et al. and emphasize the importance of distinguishing between distraction and office set-up.

\section{Discussion}
\label{sec:discussion}

\subsection{Implications and recommendations}

The COVID-19 pandemic and the subsequent lockdown have been a major professional change for many software engineers. In the present research, we investigated how a range of relevant variables are associated with and predict software engineers' well-being and productivity.
The first significant outcome of this research is that many variables are associated with well-being and productivity. The strength of the association ranges from small to large~\cite{cohen1992power}.
Also, well-being and productivity were positively associated. This implies that neglecting people's well-being might also negatively impact productivity. 
Together, our findings support Ralph et al.'s~\cite{Ralph2020pandemic} recommendation that pressuring employees to keep the average productivity level without taking care of their well-being will likely lower productivity.
However, we would also like to present an alternative interpretation that having productive employees will strengthen their sense of achievement and improve their well-being. This alternative interpretation follows from that we did not find any causal relations. This is partly driven by most variables' high stability over time, which leaves little variance to be explained by any other variable. However, it can also imply that many variables influence each other, such as well-being and productivity. Further, some of our predictors can likely be hierarchically organized. For example, introversion can lead to loneliness, resulting in more anxiety, which can cause lower levels of well-being. It will be interesting for future research to develop hierarchical models of emotions and other variables we used as predictors. This would further improve our understanding of the predictors of well-being and productivity.
Since we started this investigation only after the pandemic, we could not contrast our results with non-remote pre-pandemic settings. 
Instead, we are providing evidence-based findings to help software engineers and organizations to work remotely. 

In the following, we focus on practical recommendations based on the most reliable predictors of well-being and productivity that we identified through our longitudinal design: the need for autonomy, stress, daily routines, social contacts, need for competence, extraversion, and quality of sleep as predictors of well-being, in Table~\ref{tab:comparisonWB}. Distractions and boredom related to productivity are discussed in Table~\ref{tab:comparisonPR}.

Persistent high-stress levels are related to adverse outcomes in the workplace~\cite{bazarko2013impact} and people's well-being. 
To reduce stress, it could be helpful for some people to practice mindfulness-based stress reduction training and practices as Bazarko et al.~\cite{bazarko2013impact} recommend. They can be performed at home, and participating in such a program can lead to lower stress levels and a lower risk of work burnout. Grossman et al. recommended other stress reduction methods~\cite{grossman2004mindfulness}. Moreover, Naik et al.~\cite{naik2018effect}, who found that mindfulness meditation practices, slow breathing exercises, mindful awareness during yoga postures, and mindfulness during stressful situations and social interactions can reduce stress levels. Together, the results of these studies suggest that mindfulness practices, even when performed at home, can reduce stress, which could also improve software engineers' well-being while being quarantined. While mindfulness practices seem to be effective methods to impact peoples' well-being positively; they might not work for everyone. For some individuals, getting physically active by exercising or going for a run, taking time to disconnect and reading a book, letting loose while dancing, or even getting creative and paint might have the same or a similar effect. For example, our exploratory analysis revealed that the coping strategy self-distraction (\textit{e.g.}, reading or watching a movie to unwind from work) was more frequently used by women, which is in line with the literature~\cite{solomon2005terror}. This indicates that self-distraction as a coping strategy is more effective for women than men. So, more research is needed to find out adequate coping strategies also for men.

As part of the overall quality of life, the quality of social contacts has a significant impact on people's well-being. Therefore, employers should be interested in enabling their employees to spend time with people they value and encourage them to build strong, meaningful relationships within their work environment. Creating a virtual office (\textit{e.g.}, using an online working environment such as `Wurkr') allows people to work with the impression of sharing a physical workspace online to communicate more comfortably and work together from anywhere. 
For example, to simplify conversations, the Slack plugin `Donut'~\cite{slack2020} randomly connects employees for coffee breaks to get to know each other better by spending some time chatting virtually. 
Besides, our finding that quality of social contact, but not living alone is associated with well-being, is in line with the literature. Quality of contact with one's partner and family independently predicted depression negatively, whereas the frequency of these contact did not~\cite{teo2013social}. Together, this suggests that findings from the literature can overall be generalized to people being quarantined.

Organizing the day in a structured way at home appears to be beneficial for software professionals' well-being.
People tend to overwork when working remotely~\cite{buffer2020}. 
This could be further magnified during quarantine where usual daily routines are disrupted, and thus working might become the only meaningful activity to do.
Therefore, it is essential to develop new daily routines not to be entirely absorbed by work and prevent burnout~\cite{brooks2020}.
Hence, scheduling meetings and designating time specifically for hobbies or spending time with family and friends is helpful while working from home and helps to satisfy employees' needs for social contacts.

To fulfill people's need for autonomy, it is necessary to allow employees to act on their values and interests~\cite{wang_can_2016}. While coordinating collaborative workflows and managing projects remotely comes with its challenges~\cite{buffer2020}. It is crucial for remote workers to have flexibility in how they structure, organize, and perform their tasks~\cite{wang_can_2016}. It is, therefore, helpful to delegate work packages instead of individual tasks. This makes it easier for individuals to work self-directedly and thus to fulfill their need for autonomy.

To fulfill employees' need for competence, it is necessary to provide them with the opportunity to grow personally and advance their skill set~\cite{legault2006high}. Two of the mainly required and highly demanded skills in remote work environments are communication skills and the ability to use virtual tools, such as presentation tools or collaborative project planning tools~\cite{buffer2020}. Raising awareness of the unique requirements of virtual communication is crucial for a smooth working process. 
Thus, working remotely requires specific communication skills, such as mindful listening~\cite{mcmanus2006transparency} or asynchronous communication, which allows people to work more efficiently~\cite{jarvela2002web}. 
Collaborative tools such as GitHub, Trello, Jira, Google Docs, Klaxoon, Mural, or Slack can simplify work processes and enable interactive workflows. Besides the training and development of employees' specific virtual skill set, it is also recommended to invest in employees' personal development within the company. Taking action and offering employees the opportunity to grow will evolve their role and strengthen their loyalty towards the employer and, therefore, employee retention~\cite{kossivi2016study}.

Introverted software professionals seem to be more negatively affected by the lockdown than their more extraverted peers.
This finding is counter-intuitive since extraverted people prefer more direct contacts than introverted people~\cite{ludvigh1974extraversion}.
Our interpretation of these results is that it is even more challenging for introverts to reach out to colleagues and friends when contact opportunities are more limited.
This is because being introverted does not mean that there is no need for social contacts at all.
While in the office, they had opportunities to be involved with colleagues both in a structured or unstructured fashion, at home, it is much more difficult as they have to be more proactive to reach out to colleagues in a more formalized setting, such as online collaboration platform (\textit{e.g.}, MS Teams).
Accordingly, software organizations should regularly organize both formal and informal online meeting occasions, where introverted software engineers feel a lower entry barrier to participate.

Quality of sleep is also a relevant predictor for well-being.
Although it might sound obvious, there is a robust association between sleep, well-being, and mindfulness~\cite{howell2008sleep}.
In particular, Howell et al. found that mindfulness predicts sleep quality, and quality of sleep and mindfulness predict well-being. 

Distractions at home are a challenging obstacle to overcome while working remotely.
Designating a specific work area in the home and communicating non-disturbing times with other household members are easy and quick first steps to minimize distractions at the workplace at home. Another obstacle that distracts remote workers more frequently is cyberslacking, which is understood as spending time on the internet for non-work-related reasons during working hours~\cite{CD}. Cyberslacking and its contribution to distractions at home for remote workers were not included in this study but would be worth exploring in future research.

When people experience, boredom it makes them feel ``\textit{[...] unchallenged while they think that the situation and their actions are meaningless}''~\cite[p. 181]{van2012boredom}. Especially people who thrive in a social setting at work are in danger of being bored quickly while working in isolation from their homes. The enumerated recommendations above, such as assigning interesting, personally tailored, and challenging work packages, using collaborative tools to hold yourself accountable, and having social interactions while working remotely, also help reduce boredom at work. Ideally, employees are intrinsically motivated and feel fulfilled by what they do. If this is not the case over a more extended period, and the experienced boredom is not a negative side effect of being overwhelmed while being quarantined, it might be reasonable to discuss a new field of action and area of responsibility with the employee.

To conclude, working from home certainly comes with its challenges, of which we have addressed several in this study.
However, at least software engineers appear to adapt to the lockdown over time, as people's well-being increased, and their social contacts' perceived quality improved. 
Similar results have also been confirmed by a survey study of $2,595$ New Zealanders' remote workers~\cite{Walton2020NZadaptation}.
Walton et al. found that productivity was similar or higher than pre-lockdown, and $89$\% of professionals would like to continue to work from home, at least one day per month.
This study also reveals that the most critical challenges were switching off, collaborating with colleagues, and setting up a home office.
On the other hand, working from home led to a drastic saving of time otherwise allocated to daily commuting, a higher degree of flexibility, and increased savings. A range of further recommendations of effective self-guided interventions to tackle anxiety, depression, and stress, are summarized by Fischer et al.~\cite{Fischer2020rapid}.

\begin{table}[!ht]
\centering
\caption{Summary of key findings \& recommendations for Well-Being}
\label{tab:comparisonWB}
\begin{tabular}{@{}m{2cm}m{4.5cm}m{4.5cm}@{}}
\toprule
    & \textbf{Findings} & \textbf{Recommended Actions} \\
    \midrule
 
     Autonomy & Significant positive predictor in wave 2 ($B_{W2}= .347$). & Organizations should trust their software engineers about how to reach agreed goals, leaving them a high degree of freedom about how to schedule the day which can result in higher performance~\cite{anand2012job}. \\ \addlinespace

    \addlinespace   

    Stress & Significant negative predictor in both waves ($B_{W1}= -.605$, $B_{W2}= -.337$). & Practice mindfulness-based stress reduction training such as meditation \cite{bazarko2013impact}, yoga, sport and the Wim Hof breathing method \cite{grossman2004mindfulness}. Women are better to self-distracting themselves and coping with stress compared to men. \\ \addlinespace

    \addlinespace

    Daily routines & Significant positive predictor in wave 1 ($B_{W1}= .125$). & Establish new routines, dedicating time to work, individual hobbies, and social contacts. \\ \addlinespace
    
    \addlinespace

    Social contacts & Significant positive predictor in both waves ($B_{W1}= .224$, $B_{W2}= .304$). & Support at a company level occasions for informal meetings (\textit{e.g.}, online coffee breaks) during working hours. \cite{owl_labs_2019} \\ \addlinespace
    
    \addlinespace

    Competence & 
   Significant positive associations between competence and well-being in both waves.
    & Companies train software engineers to work in a remote setting. Similarly, software engineers should choose which kind of competencies and training they think to help their careers. \\ \addlinespace
    
    \addlinespace

    Extraversion & Positive predictor in wave 1 ($B_{W1}= .223$) & Organizations and peers should proactively reach out to introverted software engineers by involving them in work or non-work-related activities (based on our findings). \\ \addlinespace

    \addlinespace

    Quality of sleep & Significant positive predictor in wave 2 ($B_{W2}= .144$) & Schedule enough sleeping time per night and practice mindfulness for sleep transition \cite{pilcher1997sleep}.\\ \addlinespace

    \bottomrule

\end{tabular}
\end{table}

\begin{table}[!ht]
\centering
\caption{Summary of significant key findings \& recommendations for Productivity}
\label{tab:comparisonPR}
\begin{tabular}{@{}m{2cm}m{4.5cm}m{4.5cm}@{}}
\toprule
      & \textbf{Findings} & \textbf{Recommended Actions} \\
    \midrule

    Boredom & Negative predictor in wave 1 ($B_{W1}= -.053$).  & Organizations should redesign employees goals by letting them choose tasks as much as possible and diversify activities \cite{van2012boredom}.\\ \addlinespace

    \addlinespace
    
    Distractions &  Negative predictor in both waves ($B_{W1}= -.065$, $B_{W2}= -.077$) & Organizations should support software engineers to set up a dedicate home office. Routines and agreements with family members about working times also help to be more focused \cite{cardenas2004exploring,mark2018effects}. \\ \addlinespace

    \bottomrule

\end{tabular}
\end{table}

\subsection{Threats to validity}

Limitations are discussed using Gren's five-facets framework~\cite{gren2018standards}. 

\textit{Reliability}. This study used a two-wave longitudinal study, where $96$\% of the initial participants, identified through a multi-stage selection process, also participated in the second wave. Further, the test-retest reliabilities were high, and the internal consistencies (Cronbach's $ \alpha$) ranged from satisfactory to very good.

\textit{Construct validity}. 
We identified $51$ variables drawn from the literature, and a suitable measurement instrument measured each.
Where possible, we used validated instruments. 
Otherwise, we developed and reported the instruments used.
To measure the construct validity, we also reported the Cronbach's alpha of all variables across both waves.
Regarding the two dependent variables, we used a validated scale for well-being and developed a new one for productivity. We made this choice since it related well to the lockdown environment our participants were facing. Thus, we chose the Satisfaction with Life Scale for well-being, and productivity was operationalized as a proportion of time spent working and efficiency per hour, compared to the estimated regular productivity without the pandemic.
However, we note that despite many variables in our study, we still might have missed one or more relevant variables, which would have been relevant to our analysis.

\textit{Conclusion validity}.
To draw our conclusions, we used multiple statistical analyses such as correlations, between-subject $t$-tests, multiple linear regressions, and structural equation modeling.
To ensure reliable conclusions, we used conservative thresholds to reduce the risk of false-positive results. The thresholds depended on the number of comparisons for each test. 
Additionally, we did not include covariates, nor did we stop the data collection based on the results, or performed any other practice associated with increasing the likelihood of finding a positive result and increasing the probability of false-positive results~\cite{simmons2011false}.  
However, we could not make any causal conclusion since all $20$ SEM analyses provided non-significant results, using a threshold of significance that reduces the risk of false-positive findings.
Also, we have not measured participants' perception of the severity of the lockdown measures. Thus, we cannot test whether they moderate the associations we found. However, it is unlikely they would have impacted our findings, as depression and worries were found to be only weakly associated with perceptions of how the government and public reacted to the lockdown measures in spring 2020~\cite{fetzer2020global}. Further, we do not have sufficient participants from different countries in our sample to test whether objective government responses (i.e., the strictness of the lockdown~\cite{hale2020variation}) moderates the associations we found. With our data, we can only provide indirect evidence that this is unlikely to be the case: When comparing participants from the UK and USA -- the lockdown was stricter in the UK by the time we collected the data~\cite{hale2020variation} -- we found little between-country mean differences. Nevertheless, we acknowledge that this is an open research question that we cannot fully answer with our data.
Finally, we made both raw data and R analysis code openly available on Zenodo.

\textit{Internal validity}.
This study did not lead to any causal conclusion, which was the present study's primary aim.
We can not say that the analyzed variables influence well-being or productivity or vice versa.
We are also aware that our study relies on self-reports, limiting the study's validity.
Further, we adjusted some measures (\textit{e.g.}, productivity). Participants were not supposed to report their perceived productivity but to make a comparison, which has been computed independently afterward in our analysis.
We also underwent an extensive screening process, selecting over 190 software engineers of the initial $483$ initial suitable subjects, identified by a previous study of Russo \& Stol, through a multi-stage cluster sampling strategy \cite{baltes2020sampling}.
Typical problems related to longitudinal studies (\textit{e.g.}, attrition of the subjects over a long-term period) do not apply.
The dropout rate between the two waves has been low ($4$\%).
We run this study towards the end of the lockdown of the Covid-19 pandemic in the spring $2020$. 
In this way, participants were able to report rooted judgments of their conditions.
Waves were set at two weeks distance, which ensured that lockdowns had not been lifted yet during the data collection of wave 2, but was also not close enough so that variability in each of the variables would already be sufficiently high between the two-time points.
Since this was a pandemic, the surveyed countries' lockdown conditions have been similar (due to standardized WHO's recommendations).
However, we did not consider region-specific conditions (\textit{e.g.}, severity of virus spread) and recommendations.
Also, lockdown timing differed among countries.
To control these potential differences, we asked participants at each of the two waves if lockdown measures were still in place and if they were still working from home.
Since all our participants reported positively to both these conditions, we did not exclude anyone from the study.

\textit{External validity}.
An \textit{a priori} power analysis has determined our sample size.
As with any longitudinal study, we designed this study to maximize internal validity~\cite{kehr2015quantitative}.
Accordingly, we focused on finding significant effects, rather than working with a representative sample of the software engineering population (with $N \approx 500$, such as Russo and Stol~\cite{russo2020gender} did, where the research goal focused on the generalizability of results). 
Additionally, we made an effort to be able to estimate to what extent our findings depend on the current situation and whether they would also be useful to inform researchers and practitioners interested in remote work in general, beyond exceptional circumstances and potentially beyond software engineers (i.e., knowledge workers in general). 
First, we also measured participants’ previous remote work experience in the past 12 months. This was uncorrelated with well-being and productivity, indicating that the extent to which people were working remotely before the lockdown was irrelevant.
Second, we measured both generalized anxiety and Covid-19 specific anxiety. As we now clarified in the subsection “Correlations”, generalized anxiety is more relevant for people’s well-being than Covid-19 specific anxiety. This suggests that our findings are at least partly COVID-19 independent: If people were terrified by COVID-19, it would have been a stronger predictor than generalized anxiety.
Third, many of our findings relate to the findings reported in the psychological literature.
This study demonstrates that those findings also hold in a sample of professional software engineers while expanding the literature substantially through our design, including a large set of relevant variables.

\section{Conclusion}
\label{sec:conclusion}

The COVID-19 pandemic disrupted software engineers in several ways.
Abruptly, lockdown and quarantine measures changed the way of working and relating to other people.
Software engineers, in line with most knowledge workers, started to work from home with unprecedented challenges.
Most notably, our research shows that high-stress levels, the absence of daily routines, and social contacts are some of the variables most related to well-being.
Similarly, low productivity is related to boredom and distractions at home.

We base our results on a longitudinal study, which involved 192 software professionals.
After identifying $51$ relevant variables related to well-being or productivity during a quarantine from literature, we run a correlation study based on the results gathered in our first wave. 
For the second wave, we selected only the variables correlated with at least a medium effect size with well-being or productivity.
Afterward, we run $20$ structural equation modeling analyses, testing for causal relations.
We could not find any significant relation, concluding that we do not know if the dependent variables are caused by independent ones or vice versa.
Accordingly, we ran several multiple regression analyses to identify unique predictors of well-being and productivity, where we found several significant results.

This paper confirms that, on average, software engineers' well-being increased during the pandemic.
Also, there is a correlation between well-being and productivity.
Out of $51$ factors, nine were reliably associated with well-being and productivity.
Based on our findings, we proposed some actionable recommendations that might be useful to deal with potential future pandemics.

Software organizations might start to experimentally ascertain whether adopting these recommendations will increase professionals' productivity and well-being.
Our research findings indicate that granting a higher degree of autonomy to employees might be beneficial, on average. However, while extended autonomy might be perceived positively experienced by those with a high need for autonomy, it might be perceived as stressful for those who prefer structure. It is unlikely that any intervention will have the same effect on all people (since there is a substantial variation for most variables); it is essential to have individual differences in mind when exploring any interventions' effects. 
Thus, adopting incremental intervention, based on our findings, where organizations can get feedback from their employees, is recommended.

Future work will explore several directions.
Cross-sectional studies with representative samples will test whether our findings are generalizable and get a better understanding of underlying mechanisms between the variables.
We will also investigate the effectiveness of specific software tools and their effect on software engineering professionals' well-being and productivity with particular regard to the relevant variables.

\section*{Supplementary Materials} 
The full survey, raw data, and R analysis scripts are openly available under a CC BY 4.0 license on Zenodo, DOI: \url{https://doi.org/10.5281/zenodo.3959131}.

\section*{Acknowledgment} 
We thank the Editors-in-Chief for fast-tracking our manuscript and the anonymous reviewers for the supportive constructive feedback. 
The authors would also like to thank Gabriel Lins de Holanda Coelho for initial feedback on this project. 
This work was supported, in part, by the Carlsberg Foundation under grant agreement number CF20-0322 (PanTra --- Pandemic Transformation).

\bibliographystyle{spmpsci}   
\bibliography{main}  

\clearpage 
\appendix
\section{Appendix}

\begin{figure}[h]
\centering
\includegraphics[width=1\textwidth]{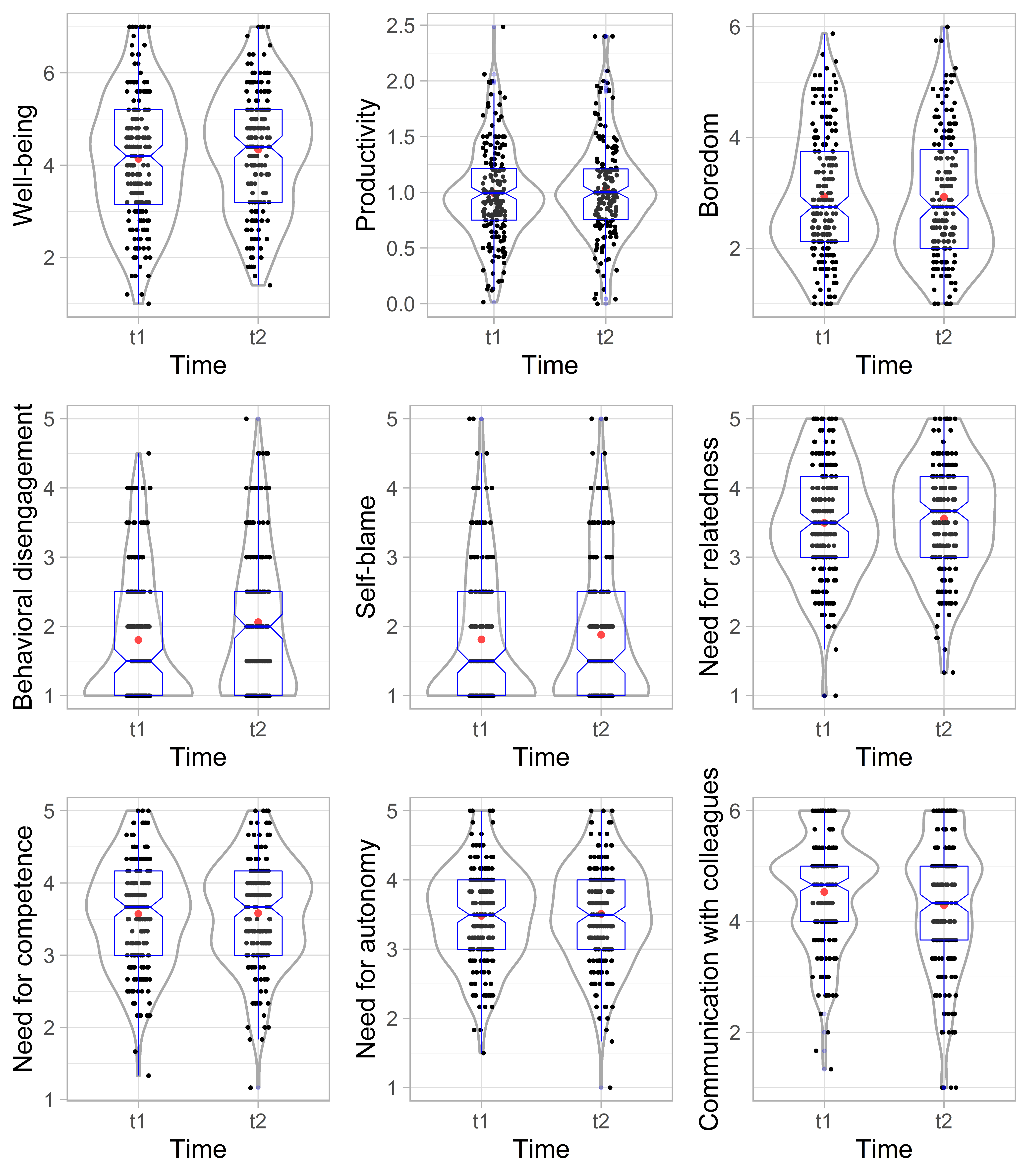}
\caption{Distributions of within-subject comparisons at time 1}
\label{fig:within1}
\end{figure}

\begin{figure}[h]
\centering
\includegraphics[width=1\textwidth]{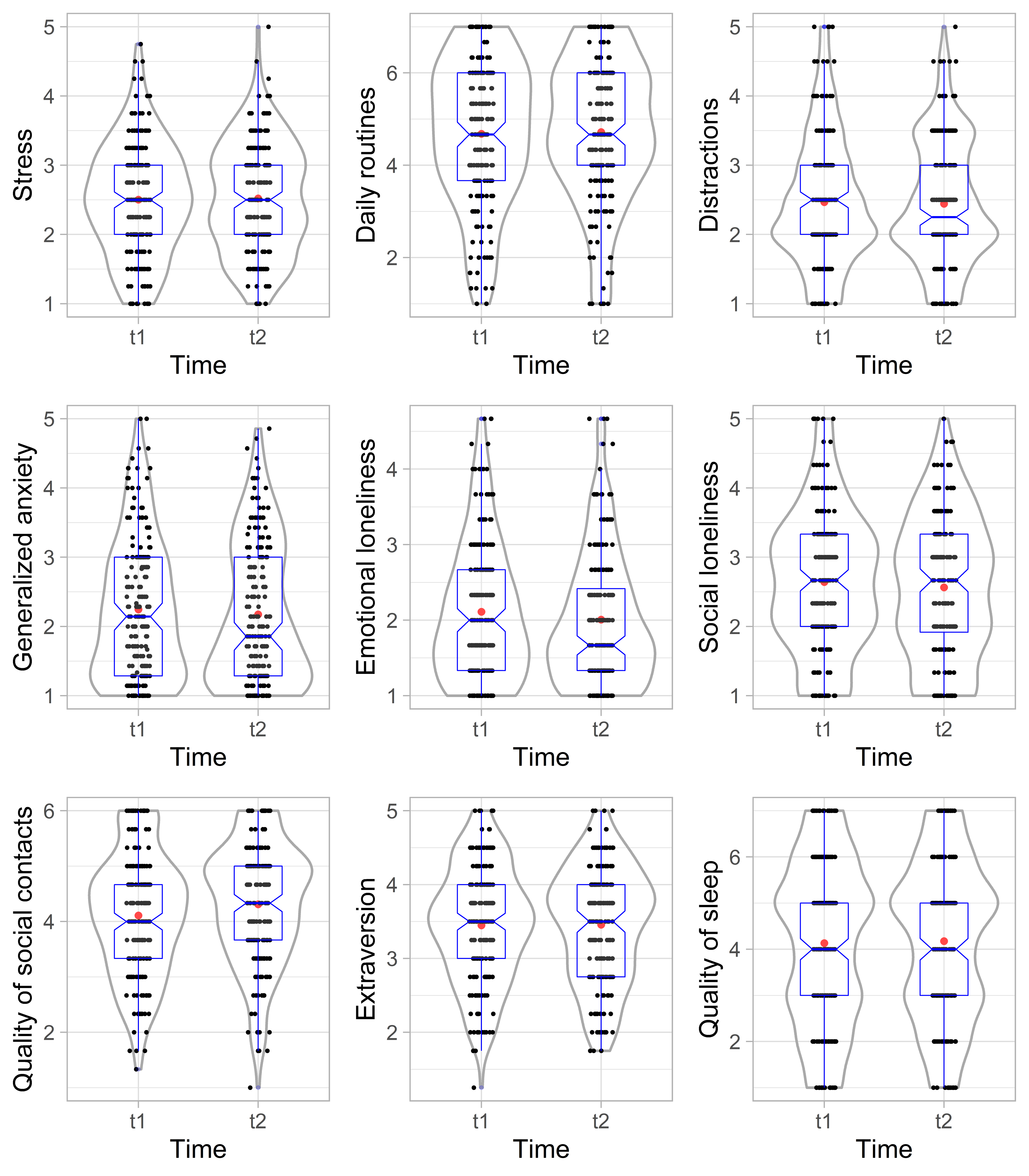}
\caption{Distributions of within-subject comparisons at wave 2}
\label{fig:within2}
\end{figure}

\begin{figure}[h]
\centering
\includegraphics[width=1\textwidth]{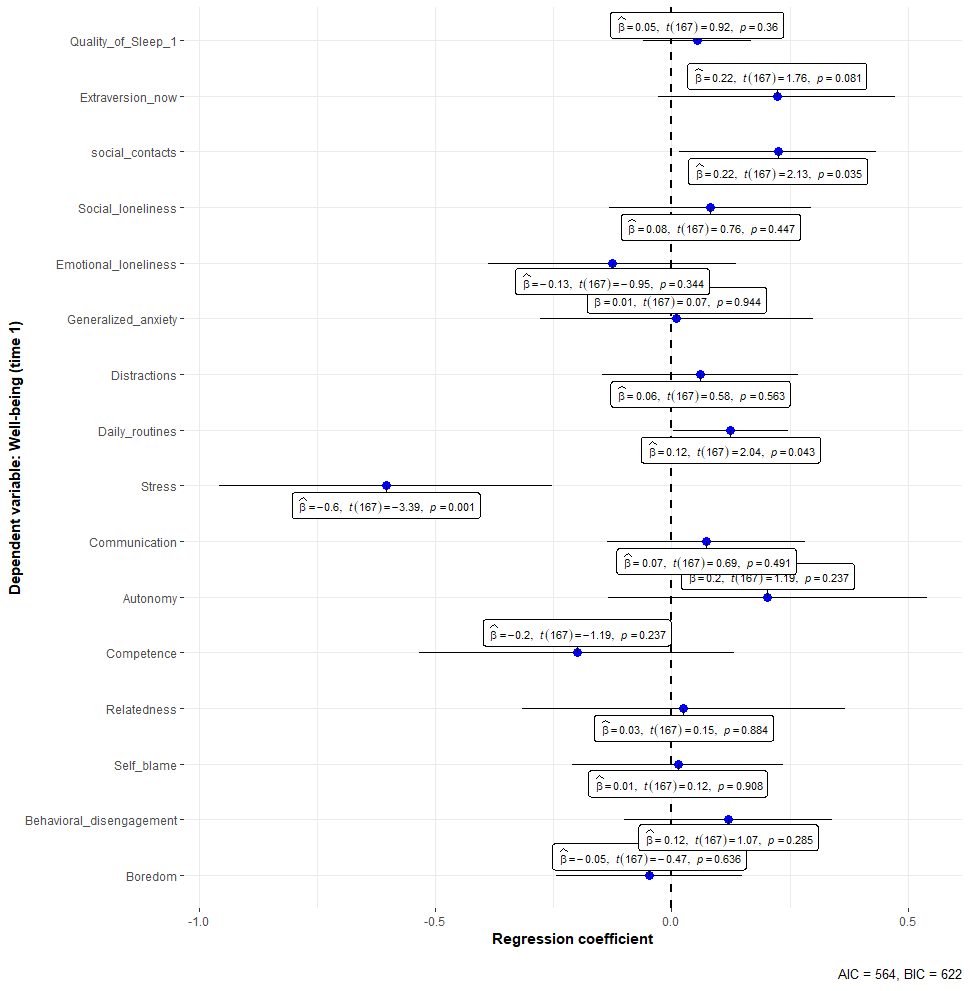}
\caption{Regression coefficients of variables predicting well-being at time 1}
\label{fig:reg_well_being1}
\end{figure}

\begin{figure}[h]
\centering
\includegraphics[width=1\textwidth]{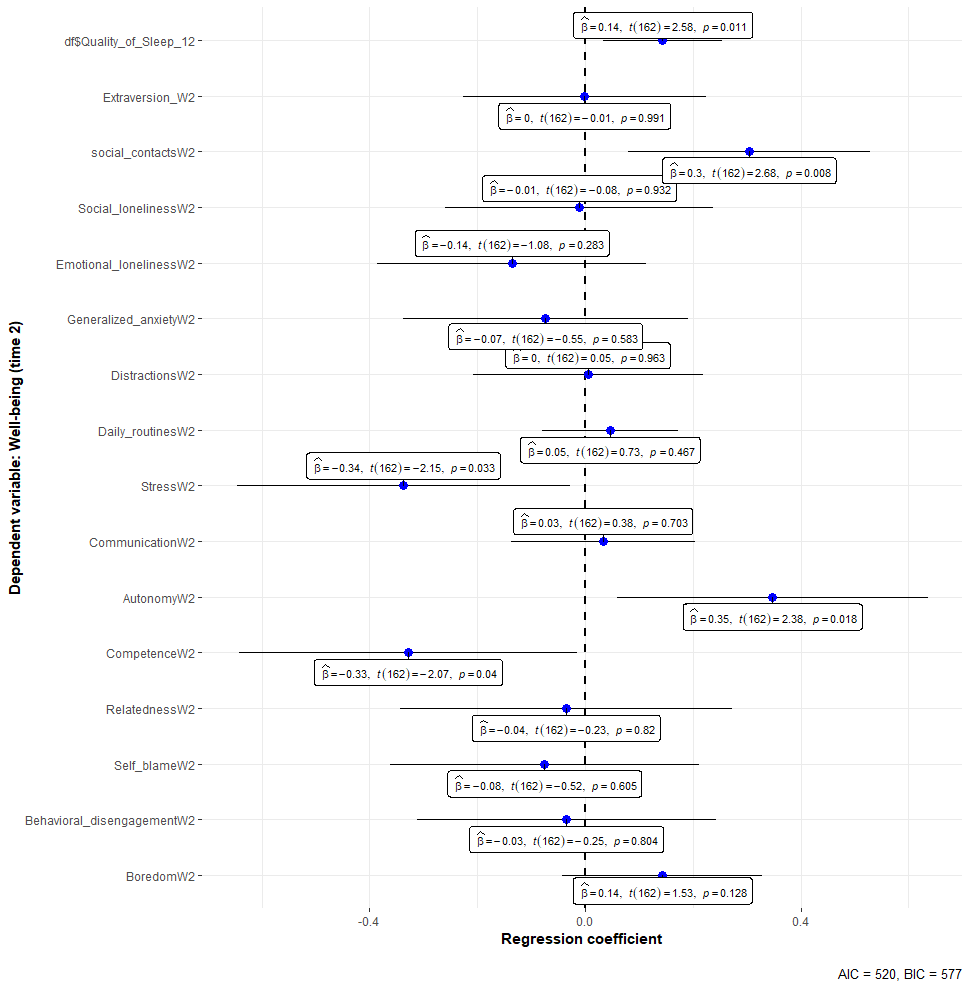}
\caption{Regression coefficients of variables predicting well-being at time 2}
\label{fig:reg_well_being2}
\end{figure}

\begin{figure}[h]
\centering
\includegraphics[width=1\textwidth]{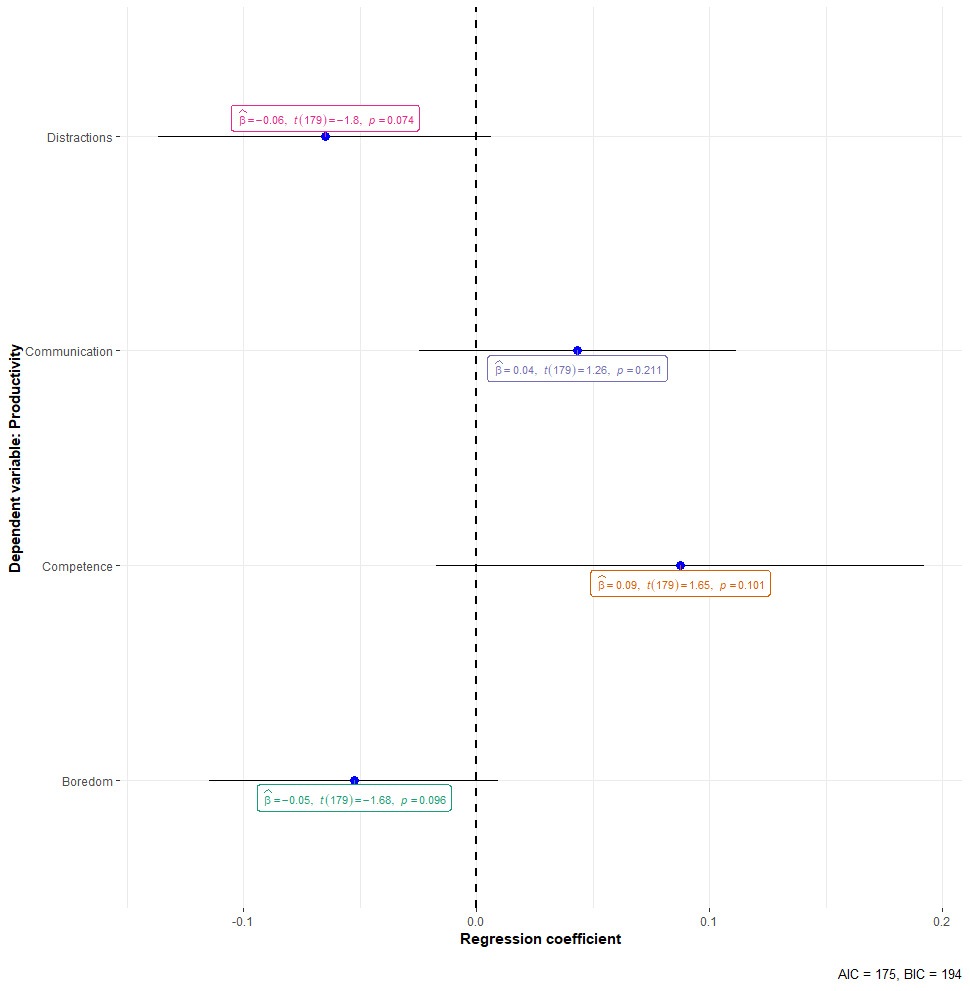}
\caption{Regression coefficients of variables predicting productivity at time 1}
\label{fig:reg_productivity1}
\end{figure}

\begin{figure}[h]
\centering
\includegraphics[width=1\textwidth]{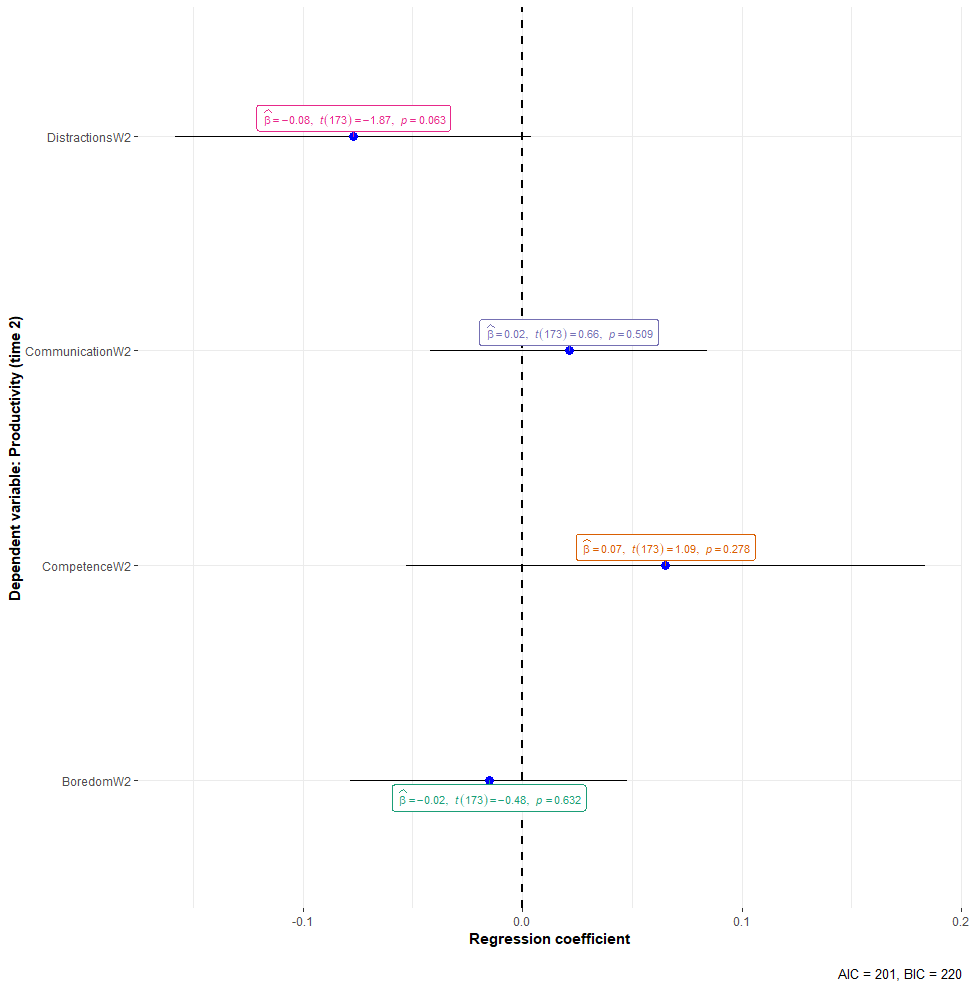}
\caption{Regression coefficients of variables predicting productivity at time 2}
\label{fig:reg_productivity2}
\end{figure}

\begin{table}[h]
\caption{Summary of constructs, instruments, reliability, and changes to instruments}

\label{tab:constructs}
\newline
\tiny \textit{Note}. NA: For some measures it is not possible to compute Cronbach's $\alpha$ (\textit{e.g.}, for one item measures). Test-retest correlation, another measure of reliability, can be found in Table~\ref{tab:correlations}. \\
\end{table}

\begin{table}[h]
\caption{Structural Equation Modeling analyses}
\sisetup{
    group-digits=true,
    detect-weight=true,
    detect-shape=true,
    table-format=-1.3,
    table-alignment = left
}
\resizebox{\textwidth}{!}{\begin{tabular}{@{}lSS[table-format=1.3]SSS[table-format=1.3]SSS[table-format=1.3]S[table-format=1.3]@{}}
\hline
                                      & \multicolumn{3}{c}{\textbf{IV$_1$ {--} DV$_1$}}     & \multicolumn{3}{c}{\textbf{IV$_2$ {--} DV$_2$}}     & \multicolumn{3}{c}{\textbf{IV$_1$ $\rightarrow$ DV$_2$}} \\
Independent variable (IV) - Dependent variable (DV)                              & \textit{B} & \textit{SE} & \textit{p}      & \textit{B} & \textit{SE} & \textit{p}      & \textit{B}      & \textit{SE}     & \textit{p}     \\ \hline
Well-being – Productivity             & 0.127      & 0.048       & 0.009           & 0.062      & 0.027       & 0.024           & 0.001           & 0.020            & 0.968          \\
Boredom – Well-being                  & -0.729     & 0.155       & {\textless} 0.001 & 0.023      & 0.064       & 0.720            & 0.011           & 0.075           & 0.880           \\
Behavioral-disengagement – Well-being & -0.484     & 0.126       & {\textless} 0.001 & -0.158     & 0.059       & 0.007           & -0.013          & 0.100             & 0.898          \\
Self-blame – Well-being               & -0.629     & 0.147       & {\textless} 0.001 & -0.167     & 0.049       & 0.001           & 0.072           & 0.088           & 0.416          \\
Distractions – Well-being             & -0.342     & 0.117       & 0.004           & -0.107     & 0.051       & 0.036           & 0.015           & 0.118           & 0.900            \\
Generalized anxiety – Well-being      & -0.698     & 0.137       & {\textless} 0.001 & -0.187     & 0.050        & {\textless} 0.001 & -0.020           & 0.086           & 0.816          \\
\hline
\end{tabular}}
\label{tab:sem}

\vspace{0.5cm}


\resizebox{\textwidth}{!}{\begin{tabular}{@{}lSS[table-format=1.3]SSS[table-format=1.3]SSS[table-format=1.3]S[table-format=1.3]@{}}\toprule
                                      & \multicolumn{3}{c}{\textbf{DV$_1$ $\rightarrow$ DV$_2$}} & \multicolumn{3}{c}{\textbf{IV$_1$ $\rightarrow$ IV$_2$}} & \multicolumn{3}{c}{\textbf{DV$_1$ $\rightarrow$ IV$_2$}} \\
Independent variable (IV) - Dependent variable (DV)                               & \textit{B}    & \textit{SE}    & \textit{p}        & \textit{B}    & \textit{SE}    & \textit{p}        & \textit{B}      & \textit{SE}     & \textit{p}     \\ \midrule
Well-being {–} Productivity             & 0.529         & 0.070           & {\textless}0.001   & 0.698         & 0.067          & {\textless}0.001   & 0.112           & 0.179           & 0.531          \\
Boredom {–} Well-being                  & 0.711         & 0.072          & {\textless}0.001   & 0.753         & 0.096          & {\textless}0.001   & -0.076          & 0.054           & 0.156          \\
Behavioral-disengagement {–} Well-being & 0.700           & 0.070           & {\textless}0.001   & 0.632         & 0.101          & {\textless}0.001   & -0.092          & 0.046           & 0.046          \\
Self-blame {–} Well-being               & 0.724         & 0.073          & {\textless}0.001   & 0.533         & 0.088          & {\textless}0.001   & -0.041          & 0.037           & 0.276          \\
Distractions {–} Well-being             & 0.705         & 0.069          & {\textless}0.001   & 0.827         & 0.133          & {\textless}0.001   & -0.056          & 0.041           & 0.17           \\
Generalized anxiety {–} Well-being      & 0.711         & 0.075          & {\textless}0.001   & 0.730          & 0.072          & {\textless}0.001   & -0.100            & 0.010            & 0.011          \\
\bottomrule

\end{tabular}}
\tiny \textit{Note}. B: unstandardized regression estimate, SE: Standard Error, $p$: $p-value$, CFI: Comparative Fit Index, RMSEA: Root Mean Square Error of Approximation, SRMR: Standardized Root Mean Square Residual. Variable$_1$ is measured in the first wave, and Variable$_2$ in the second wave. \\
\end{table}

\begin{table}[]
\sisetup{
    group-digits=true,
    detect-weight=true,
    detect-shape=true,
    table-format=-1.3,
    table-alignment = left
}
\caption{Within-subject comparisons (mixed-effects model): Are there any mean changes over time?}
\label{tab:WithinComparisons}
\resizebox{\textwidth}{!}{%
\begin{tabular}{@{}lllllSllSlll@{}}
\toprule
                             & \multicolumn{2}{l}{Time 1} & \multicolumn{2}{l}{Time 2} &       &      &               &                 &        &         &       \\
                             & M            & SD          & M            & SD          & B     & SE   & {95\%-CI}       & \textit{p}      & Higher & Smaller & Equal \\ \midrule
Well-being                   & 4.140        & 1.367       & 4.340        & 1.289       & .180  & .074 & .035,   .325  & .015            & 91     & 70      & 23    \\
Productivity                 & 0.990        & 0.419       & 1.032        & 0.436       & .046  & .031 & -.016,  .107  & .147            & 87     & 77      & 19    \\
Boredom                      & 2.936        & 1.136       & 2.927        & 1.158       & .013  & .066 & -0.117, 0.142 & .848            & 91     & 79      & 14    \\
Behavioral-disengagement     & 1.805        & 0.936       & 2.062        & 1.030       & .255  & .069 & .119, .391    & \textless{}.001 & 82     & 40      & 62    \\
Self-blame                   & 1.812        & 0.990       & 1.880        & 1.013       & .065  & .065 & -.063, .192   & .319            & 60     & 52      & 72    \\
Need for Relatedness         & 3.497        & 0.830       & 3.559        & 0.803       & .054  & .046 & -.035, .144   & .233            & 86     & 73      & 25    \\
Need for Competence          & 3.572        & 0.735       & 3.582        & 0.731       & .005  & .045 & -.084  .094   & .915            & 82     & 82      & 20    \\
Need for Autonomy            & 3.483        & 0.688       & 3.511        & 0.732       & .023  & .036 & -.049, .094   & .535            & 88     & 67      & 29    \\
Communication                & 4.534        & 0.996       & 4.292        & 1.185       & -.223 & .067 & -.360 -.095   & \textless{}.001 & 57     & 81      & 38    \\
Stress                       & 2.501        & 0.807       & 2.520        & 0.797       & .024  & .043 & -.061, .109   & .581            & 81     & 64      & 39    \\
Daily routines               & 4.681        & 1.561       & 4.717        & 1.533       & .017  & .083 & -.147, .180   & .843            & 71     & 72      & 41    \\
Distractions                 & 2.466        & 0.934       & 2.443        & 0.895       & -.015 & .058 & -.128, .098   & .789            & 58     & 64      & 62    \\
Generalized anxiety          & 2.245        & 1.000       & 2.174        & 1.010       & -.065 & .051 & -.166, .035   & .202            & 69     & 90      & 25    \\
Emotional loneliness         & 2.111        & 0.903       & 2.007        & 0.871       & -.102 & .049 & -.198, -.006  & .037            & 54     & 79      & 51    \\
Social loneliness            & 2.641        & 1.004       & 2.563        & 1.017       & -.056 & .058 & -.171, .058   & .334            & 65     & 79      & 40    \\
Quality of social   contacts & 4.109        & 1.093       & 4.312        & 1.077       & .182  & .066 & .053, .312    & .006            & 91     & 54      & 39    \\
Extraversion                 & 3.448        & 0.786       & 3.457        & 0.778       & .007  & .035 & -.061, .076   & .838            & 73     & 59      & 52    \\
Quality of Sleep             & 4.130        & 1.754       & 4.174        & 1.686       & -.010 & .088 & -.182, .162   & .906            & 54     & 51      & 79    \\ \bottomrule
\end{tabular}%
}
\small \textit{Note}. B: fixed effects estimates, SE: standard error, CI: confidence interval; Higher: Absolute number of people who scored higher on a variable at time 2 compared to time 1; Lower: Number of people who scored lower at time 2; Equal: People whose score has not changed over time. \\
\end{table}

\begin{table}[]
\sisetup{
    group-digits=true,
    detect-weight=true,
    detect-shape=true,
    table-format=-1.3,
    table-alignment = left
}
\caption{Comparison of women and men}
\resizebox{\textwidth}{!}{
\label{tab:womenmen}
\begin{tabular}{@{}lS[table-format=2.3]lS[table-format=2.3]lSlSS[table-format=1.4]@{}}
\toprule
                                    & \textbf{Men} & \textbf{}   & \textbf{Women} & \textbf{}   & \textbf{}           & \textbf{}           & \textbf{}  & \textbf{}    \\
                                    & \textbf{M}   & \textbf{SD} & \textbf{M}     & \textbf{SD} & \textit{\textbf{t}} & \textit{\textbf{p}} & \textbf{d} & \textbf{OVL} \\ \midrule
Well-being                & 4.109  & 1.336  & 4.263  & 1.495  & -0.581 & 0.5639 & -0.113 & 95.494 \\
Well-being$_{t2}$               & 4.388  & 1.258  & 4.151  & 1.407  & 0.933  & 0.3553 & 0.183  & 92.710  \\
Productivity              & 1.008  & 0.416  & 0.917  & 0.430   & 1.180   & 0.2430  & 0.218  & 91.320  \\
Productivity$_{t2}$             & 1.029  & 0.433  & 1.043  & 0.452  & -0.175 & 0.8620  & -0.033 & 98.684 \\
Boredom                   & 2.942  & 1.130   & 2.908  & 1.179  & 0.163  & 0.8713 & 0.030   & 98.803 \\
Boredom$_{t2}$                   & 2.923  & 1.122  & 2.943  & 1.309  & -0.082 & 0.9353 & -0.016 & 99.362 \\
Behavioral-disengagement  & 1.799  & 0.935  & 1.829  & 0.953  & -0.176 & 0.8611 & -0.032 & 98.723 \\
Behavioral-disengagement$_{t2}$ & 1.997  & 0.957  & 2.324  & 1.259  & -1.479 & 0.1458 & -0.320  & 87.288 \\
Self-blame                & 1.753  & 0.957  & 2.053  & 1.095  & -1.546 & 0.1283 & -0.304 & 87.919 \\
Self-blame$_{t2}$               & 1.830   & 0.955  & 2.081  & 1.211  & -1.173 & 0.2465 & -0.248 & 90.132 \\
Relatedness               & 3.483  & 0.801  & 3.557  & 0.948  & -0.446 & 0.6577 & -0.089 & 96.451 \\
Relatedness$_{t2}$              & 3.560   & 0.748  & 3.554  & 1.004  & 0.034  & 0.9728 & 0.008  & 99.681 \\
Competence                & 3.566  & 0.704  & 3.596  & 0.862  & -0.202 & 0.8407 & -0.041 & 98.364 \\
Competence$_{t2}$               & 3.605  & 0.692  & 3.491  & 0.874  & 0.740   & 0.4628 & 0.156  & 93.783 \\
Autonomy                  & 3.476  & 0.668  & 3.509  & 0.771  & -0.239 & 0.8119 & -0.047 & 98.125 \\
Autonomy$_{t2}$                 & 3.526  & 0.698  & 3.45   & 0.863  & 0.494  & 0.6236 & 0.103  & 95.893 \\
Communication             & 4.511  & 1.004  & 4.623  & 0.972  & -0.625 & 0.5343 & -0.112 & 95.534 \\
Communication$_{t2}$            & 4.293  & 1.118  & 4.288  & 1.434  & 0.020   & 0.9839 & 0.004  & 99.840  \\
Stress                    & 2.468  & 0.744  & 2.638  & 1.025  & -0.966 & 0.3391 & -0.212 & 91.558 \\
Stress$_{t2}$                   & 2.485  & 0.748  & 2.662  & 0.965  & -1.042 & 0.3025 & -0.223 & 91.122 \\
Daily routines            & 4.758  & 1.469  & 4.368  & 1.877  & 1.191  & 0.2394 & 0.250   & 90.052 \\
Daily routines$_{t2}$           & 4.902  & 1.439  & 3.982  & 1.689  & 3.049  & 0.0037 & 0.617  & 75.770  \\
Distractions              & 2.481  & 0.884  & 2.408  & 1.126  & 0.370   & 0.7127 & 0.078  & 96.889 \\
Distractions$_{t2}$             & 2.459  & 0.869  & 2.378  & 1.003  & 0.449  & 0.6550  & 0.090   & 96.411 \\
Generalized anxiety       & 2.123  & 0.916  & 2.738  & 1.175  & -3.007 & 0.0042 & -0.632 & 75.200   \\
Generalized anxiety$_{t2}$      & 2.098  & 0.963  & 2.475  & 1.144  & -1.844 & 0.0712 & -0.376 & 85.088 \\
Emotional loneliness      & 2.022  & 0.848  & 2.474  & 1.033  & -2.498 & 0.0158 & -0.510  & 79.872 \\
Emotional loneliness$_{t2}$     & 1.975  & 0.837  & 2.135  & 0.998  & -0.899 & 0.3730  & -0.184 & 92.670  \\
Social loneliness         & 2.667  & 0.969  & 2.535  & 1.146  & 0.653  & 0.5169 & 0.131  & 94.778 \\
Social loneliness$_{t2}$        & 2.560   & 0.985  & 2.577  & 1.151  & -0.080  & 0.9365 & -0.016 & 99.362 \\
Social contacts           & 4.032  & 1.068  & 4.421  & 1.149  & -1.893 & 0.0637 & -0.358 & 85.794 \\
Social contacts$_{t2}$          & 4.281  & 1.055  & 4.432  & 1.165  & -0.719 & 0.4754 & -0.140  & 94.419 \\
Extraversion              & 3.401  & 0.786  & 3.638  & 0.766  & -1.700   & 0.0944 & -0.303 & 87.958 \\
Extraversion$_{t2}$             & 3.405  & 0.775  & 3.662  & 0.769  & -1.817 & 0.0745 & -0.333 & 86.776 \\
Quality of sleep          & 4.221  & 1.746  & 3.763  & 1.762  & 1.436  & 0.1564 & 0.262  & 89.578 \\
Quality of sleep$_{t2}$         & 4.299  & 1.673  & 3.676  & 1.668  & 2.032  & 0.0469 & 0.373  & 85.205 \\
Compliance                & 7.190   & 1.093  & 7.518  & 0.700    & -2.278 & 0.0251 & -0.319 & 87.328 \\
Belief in conspiracies    & 2.379  & 1.244  & 2.316  & 1.205  & 0.289  & 0.7738 & 0.051  & 97.966 \\
Self-distraction          & 3.182  & 1.040   & 3.750   & 0.86   & -3.491 & 0.0009 & -0.564 & 77.794 \\
Active coping             & 3.188  & 0.922  & 3.447  & 0.884  & -1.604 & 0.1140  & -0.283 & 88.748 \\
Denial                    & 1.425  & 0.725  & 1.355  & 0.603  & 0.615  & 0.5407 & 0.100    & 96.012 \\
Substance use             & 1.682  & 1.087  & 1.855  & 1.330   & -0.745 & 0.4599 & -0.152 & 93.942 \\
Emotional support         & 2.643  & 1.032  & 3.039  & 1.042  & -2.105 & 0.0398 & -0.384 & 84.774 \\
Instrumental support      & 2.162  & 0.967  & 2.250   & 0.957  & -0.505 & 0.6157 & -0.091 & 96.371 \\
Venting                   & 2.338  & 1.018  & 2.408  & 0.965  & -0.398 & 0.6924 & -0.070  & 97.208 \\
Positive reframing        & 3.081  & 1.019  & 3.566  & 1.067  & -2.531 & 0.0143 & -0.471 & 81.382 \\
Planning                  & 2.984  & 1.044  & 3.289  & 1.125  & -1.521 & 0.1340  & -0.288 & 88.550  \\
Humor                     & 2.503  & 1.174  & 2.855  & 1.330   & -1.494 & 0.1411 & -0.292 & 88.392 \\
Acceptance                & 3.935  & 0.768  & 4.276  & 0.742  & -2.522 & 0.0144 & -0.447 & 82.315 \\
Religion                  & 1.662  & 0.925  & 1.934  & 1.439  & -1.110  & 0.2731 & -0.260  & 89.657 \\
Office setup              & 15.729 & 1.108  & 15.474 & 1.239  & 1.163  & 0.2501 & 0.225  & 91.043 \\
Self-control              & 3.338  & 0.834  & 3.246  & 0.919  & 0.563  & 0.5760  & 0.108  & 95.694 \\
Volunteering              & 3.346  & 1.147  & 3.904  & 1.257  & -2.489 & 0.0160  & -0.477 & 81.149 \\
Diet                      & 9.916  & 3.413  & 9.618  & 3.791  & 0.441  & 0.6609 & 0.085  & 96.610  \\
Exercising overall        & 10.078 & 8.542  & 10.776 & 17.730  & -0.236 & 0.8144 & -0.064 & 97.447 \\
Financial situation       & 7.260   & 1.648  & 7.079  & 2.110   & 0.492  & 0.6247 & 0.103  & 95.893 \\
COVID-19 anxiety          & 3.244  & 1.036  & 3.395  & 1.158  & -0.736 & 0.4650  & -0.143 & 94.300   \\
Mental exercise           & 4.497  & 1.377  & 4.750   & 1.478  & -0.958 & 0.3422 & -0.181 & 92.789 \\
Extrinsic-social          & 3.591  & 1.311  & 3.763  & 1.518  & -0.643 & 0.5233 & -0.127 & 94.937 \\
Extrinsic-materialistic   & 4.264  & 1.324  & 4.623  & 1.387  & -1.441 & 0.1553 & -0.268 & 89.340  \\
Intrinsic motivation      & 4.468  & 1.421  & 5.018  & 1.427  & -2.130  & 0.0375 & -0.387 & 84.657 \\
People living in the same household & 2.221        & 2.789       & 1.579          & 1.703       & 1.802               & 0.0748              & 0.246      & 90.211       \\
Technological skills      & 27.240  & 0.776  & 27.211 & 1.069  & 0.161  & 0.8726 & 0.035  & 98.604 \\
Previous remote work experience     & 40.890        & 36.340       & 39.947         & 37.649      & 0.139               & 0.8899              & 0.026      & 98.963       \\
Children at home          & 0.714 &	1.008 &	0.421 &	0.826 &	1.871 &	0.0657 &	0.301 &	88.037 \\ \bottomrule
0\end{tabular}}
\small \textit{Note}. d: Cohen's d (standardized mean difference), OVL: Overlapping coefficient~\cite{inman1989overlapping}, which estimates the amount of similarities between women and men (OVL of 100 indicates the groups are identical, a OVL of 50 suggests that 50\% of the responses given by women are mirrored by men). \\
\end{table}

\begin{table}[]
\sisetup{
    group-digits=true,
    detect-weight=true,
    detect-shape=true,
    table-format=-1.3,
    table-alignment = left
}
\caption{Comparison between UK and USA-based participants}
\label{tab:country}
\resizebox{\textwidth}{!}{%
\begin{tabular}{@{}lS[table-format=2.3]S[table-format=2.3]S[table-format=2.3]S[table-format=2.3]SlSS[table-format=1.4]@{}}
\toprule
                                      & \multicolumn{2}{l}{UK} & \multicolumn{2}{l}{USA} &            &            &            &        \\
                                      & M          & {SD}        & M          & {SD}         & \textit{t} & \textit{p} & \textit{d} & {OVL}    \\ \midrule
Well-being                            & 4.248      & 1.302     & 4.288      & 1.448      & -0.158     & 0.8752     & -0.030      & 98.803 \\
Well-being$_{t2}$                           & 4.294      & 1.220      & 4.392      & 1.461      & -0.381     & 0.7039     & -0.074     & 97.049 \\
Productivity                          & 1.018      & 0.453     & 0.936      & 0.385      & 1.047      & 0.2975     & 0.193      & 92.312 \\
Productivity$_{t2}$                         & 0.977      & 0.414     & 1.076      & 0.472      & -1.162     & 0.2480      & -0.225     & 91.043 \\
Boredom                               & 2.857      & 1.072     & 2.889      & 1.194      & -0.151     & 0.8802     & -0.029     & 98.843 \\
Boredom$_{t2}$                              & 2.960       & 1.159     & 2.740       & 1.166      & 0.994      & 0.3226     & 0.189      & 92.471 \\
Behavioral-disengagement              & 1.865      & 0.885     & 1.683      & 0.852      & 1.123      & 0.2641     & 0.210       & 91.638 \\
Behavioral-disengagement$_{t2}$             & 2.089      & 0.952     & 1.910       & 1.024      & 0.948      & 0.3456     & 0.182      & 92.749 \\
Self-blame                            & 1.786      & 0.932     & 1.740       & 1.059      & 0.241      & 0.8100       & 0.046      & 98.165 \\
Self-blame$_{t2}$                           & 1.944      & 1.025     & 1.680       & 0.896      & 1.451      & 0.1498     & 0.272      & 89.182 \\
Relatedness                           & 3.521      & 0.772     & 3.545      & 0.827      & -0.158     & 0.8750      & -0.030      & 98.803 \\
Relatedness$_{t2}$                          & 3.538      & 0.711     & 3.593      & 0.847      & -0.372     & 0.7111     & -0.072     & 97.128 \\
Competence                            & 3.569      & 0.734     & 3.593      & 0.873      & -0.159     & 0.8742     & -0.030      & 98.803 \\
Competence$_{t2}$                           & 3.605      & 0.670      & 3.650       & 0.815      & -0.315     & 0.7534     & -0.061     & 97.567 \\
Autonomy                              & 3.503      & 0.700       & 3.516      & 0.754      & -0.098     & 0.9223     & -0.018     & 99.282 \\
Autonomy$_{t2}$                             & 3.565      & 0.751     & 3.443      & 0.818      & 0.808      & 0.4210      & 0.155      & 93.823 \\
Communication                         & 4.472      & 1.031     & 4.593      & 1.000          & -0.624     & 0.5341     & -0.119     & 95.255 \\
Communication$_{t2}$                        & 4.383      & 1.237     & 4.245      & 1.263      & 0.573      & 0.5678     & 0.110       & 95.614 \\
Stress                                & 2.528      & 0.713     & 2.312      & 0.871      & 1.430       & 0.1560      & 0.273      & 89.143 \\
Stress$_{t2}$                               & 2.573      & 0.690      & 2.340       & 0.890       & 1.516      & 0.1330      & 0.296      & 88.234 \\
Daily routines                        & 4.889      & 1.409     & 4.474      & 1.738      & 1.385      & 0.1693     & 0.265      & 89.459 \\
Daily routines$_{t2}$                       & 4.817      & 1.531     & 4.647      & 1.646      & 0.562      & 0.5752     & 0.108      & 95.694 \\
Distractions                          & 2.532      & 0.920      & 2.385      & 1.018      & 0.806      & 0.4222     & 0.152      & 93.942 \\
Distractions$_{t2}$                         & 2.540       & 0.816     & 2.500        & 0.985      & 0.232      & 0.8168     & 0.045      & 98.205 \\
Generalized anxiety                   & 2.265      & 0.942     & 2.134      & 1.075      & 0.689      & 0.4926     & 0.131      & 94.778 \\
Generalized anxiety$_{t2}$                  & 2.219      & 0.965     & 1.989      & 0.962      & 1.257      & 0.2114     & 0.239      & 90.488 \\
Emotional   loneliness                & 2.048      & 0.956     & 2.038      & 0.802      & 0.056      & 0.9555     & 0.010       & 99.601 \\
Emotional   loneliness$_{t2}$               & 2.059      & 0.938     & 1.887      & 0.796      & 1.052      & 0.2949     & 0.197      & 92.154 \\
Social loneliness                     & 2.619      & 0.912     & 2.583      & 1.074      & 0.190       & 0.8498     & 0.036      & 98.564 \\
Social   loneliness$_{t2}$                  & 2.527      & 0.942     & 2.493      & 1.135      & 0.168      & 0.8673     & 0.032      & 98.723 \\
Social contacts                       & 4.053      & 1.091     & 4.218      & 1.064      & -0.818     & 0.4150      & -0.153     & 93.902 \\
Social   contacts$_{t2}$                    & 4.274      & 1.120      & 4.327      & 1.189      & -0.238     & 0.8121     & -0.046     & 98.165 \\
Extraversion                          & 3.552      & 0.728     & 3.486      & 0.799      & 0.459      & 0.6473     & 0.087      & 96.530  \\
Extraversion$_{t2}$                         & 3.516      & 0.742     & 3.530       & 0.798      & -0.094     & 0.9250      & -0.018     & 99.282 \\
Quality of   sleep                    & 4.111      & 1.752     & 3.981      & 1.686      & 0.405      & 0.6860      & 0.076      & 96.969 \\
Quality of   sleep$_{t2}$                   & 3.968      & 1.599     & 4.440       & 1.704      & -1.499     & 0.1371     & -0.287     & 88.590  \\
Compliance                            & 7.450       & 0.672     & 7.065      & 1.291      & 1.925      & 0.0582     & 0.385      & 84.735 \\
Belief in   conspiracies              & 2.263      & 1.210      & 2.258      & 1.262      & 0.025      & 0.9801     & 0.005      & 99.801 \\
Self-distraction                      & 3.167      & 1.063     & 3.365      & 0.966      & -1.049     & 0.2962     & -0.195     & 92.233 \\
Active coping                         & 3.246      & 0.847     & 3.183      & 1.029      & 0.356      & 0.7230      & 0.068      & 97.288 \\
Denial                                & 1.365      & 0.624     & 1.404      & 0.700        & -0.310      & 0.7569     & -0.059     & 97.647 \\
Substance use                         & 1.968      & 1.167     & 1.683      & 1.221      & 1.274      & 0.2056     & 0.240       & 90.448 \\
Emotional support                     & 2.659      & 0.962     & 3.010       & 1.096      & -1.805     & 0.0740      & -0.342     & 86.422 \\
Instrumental   support                & 2.206      & 0.923     & 2.365      & 1.039      & -0.859     & 0.3924     & -0.163     & 93.504 \\
Venting                               & 2.254      & 0.962     & 2.404      & 1.039      & -0.796     & 0.4278     & -0.150      & 94.021 \\
Positive reframing                    & 3.230       & 0.897     & 2.981      & 1.142      & 1.282      & 0.2030      & 0.246      & 90.211 \\
Planning                              & 2.976      & 1.014     & 3.077      & 1.095      & -0.508     & 0.6128     & -0.096     & 96.172 \\
Humor                                 & 2.706      & 1.275     & 2.327      & 1.184      & 1.652      & 0.1013     & 0.307      & 87.800   \\
Acceptance                            & 4.032      & 0.772     & 4.019      & 0.671      & 0.093      & 0.9261     & 0.017      & 99.322 \\
Religion                              & 1.571      & 0.946     & 1.971      & 1.148      & -2.010      & 0.0471     & -0.384     & 84.774 \\
Office setup                          & 15.714     & 0.991     & 15.609     & 1.290       & 0.483      & 0.6303     & 0.093      & 96.291 \\
Self-control                          & 3.159      & 0.762     & 3.442      & 0.769      & -1.977     & 0.0506     & -0.371     & 85.284 \\
Volunteering                          & 3.413      & 1.132     & 3.615      & 1.232      & -0.911     & 0.3646     & -0.172     & 93.147 \\
Diet                                  & 10.627     & 3.217     & 9.740       & 3.448      & 1.415      & 0.1601     & 0.267      & 89.380  \\
Exercising   overall                  & 13.830      & 14.415    & 9.493      & 7.935      & 2.043      & 0.0437     & 0.363      & 85.598 \\
Financial situation                   & 7.302      & 1.950      & 7.135      & 1.858      & 0.469      & 0.6400       & 0.087      & 96.530  \\
Covid-19   anxiety                    & 3.397      & 1.013     & 3.067      & 1.129      & 1.632      & 0.1058     & 0.309      & 87.722 \\
Mental exercise                       & 4.516      & 1.459     & 4.423      & 1.330       & 0.356      & 0.7222     & 0.066      & 97.367 \\
Extrinsic-social                      & 3.476      & 1.257     & 3.564      & 1.515      & -0.334     & 0.7389     & -0.064     & 97.447 \\
Extrinsic-materialistic               & 3.915      & 1.214     & 4.776      & 1.310       & -3.623     & 0.0004     & -0.684     & 73.235 \\
Intrinsic   motivation                & 4.344      & 1.444     & 4.442      & 1.418      & -0.367     & 0.7142     & -0.069     & 97.248 \\
People living   in the same household & 1.921      & 1.559     & 1.962      & 1.441      & -0.146     & 0.8842     & -0.027     & 98.923 \\
Technological   skills                & 27.111     & 0.863     & 27.327     & 0.760       & -1.425     & 0.1569     & -0.264     & 89.498 \\
Previous remote   work experience     & 38.063     & 33.773    & 42.846     & 38.185     & -0.704     & 0.4830      & -0.133     & 94.698 \\
Children at home                          & 0.762      & 1.132     & 0.712      & 1.035      & 0.249      & 0.8039     & 0.046      & 98.165 \\ \bottomrule
\end{tabular}%
}
\small \textit{Note}. d: Cohen’s d (standardized mean difference), OVL: Overlapping coefficient~\cite{inman1989overlapping}, which estimates the amount of similarities between British and US-American participants (OVL of 100 indicates the groups are identical, a OVL of 50 suggests that 50\% of the responses given by British participants are mirrored by US-American participants). \\
\end{table}

\begin{sidewaystable}
\sisetup{
    group-digits=true,
    detect-weight=true,
    detect-shape=true,
    table-format=-1.3,
    table-alignment = left
}
\caption{Correlations of well-being and productivity with all predictors at time 1}
\resizebox{17cm}{!}{
\label{tab:cor1_complete}
\begin{tabular}{@{}lSS[table-format=1.3]SSSSSS[table-format=1.3]SSSSSS[table-format=1.3]SSSSSS[table-format=1.3]SSSS@{}}
\toprule
                         & \multicolumn{6}{l}{\textbf{Well-being   (Pearson’s r)}}                                     & \multicolumn{6}{l}{\textbf{Well-being (Spearman’s   rho)}}                                     & \multicolumn{6}{l}{\textbf{Productivity   (Pearson’s r)}}                                   & \multicolumn{6}{l}{\textbf{Productivity   (Spearman’s rho)}}                                   \\
                         & \textbf{\textit{r}} & \textbf{\textit{p}} & \textbf{95LL} & \textbf{95UL} & \textbf{99.9LL} & \textbf{99.9UL} & \textbf{\textit{r\_s}} & \textbf{\textit{p}} & \textbf{95LL} & \textbf{95UL} & \textbf{99.9LL} & \textbf{99.9UL} & \textbf{\textit{r}} & \textbf{\textit{p}} & \textbf{95LL} & \textbf{95UL} & \textbf{99.9LL} & \textbf{99.9UL} & \textbf{\textit{r\_s}} & \textbf{\textit{p}} & \textbf{95LL} & \textbf{95UL} & \textbf{99.9LL} & \textbf{99.9UL} \\ \midrule
Well-being (WB)          & 1          & {-}        & {-}           & {-}           & {-}             & {-}             & 1             & {-}        & {-}           & {-}           & {-}             & {-}             & .18        & .013       & .039          & .313          & -.058           & .398            & .255          & 0          & .117          & .383          & .021            & .462            \\
Productivity (PR)        & .180        & .0130       & .039          & .313          & -.058           & .398            & .255          & 0          & .117          & .383          & .021            & .462            & 1          & {-}        & {-}           & {-}           & {-}             & {-}             & 1             & {-}        & {-}           & {-}           & {-}             & {-}             \\
\bottomrule
\end{tabular}}

\resizebox{17cm}{!}{
\textit{Note}. \textit{r}: Pearson correlation coefficient, \textit{$r_s$}: Spearman's rank correlation coefficient (Spearman's rho), \textit{p}: $p-value$ (a $p-value$ of 0 indicates $p < .001$), 95LL: lower limit of the 95\%-confidence interval, 95UL: upper limit of the 95\%-confidence interval, 99.9LL: lower limit of the 99.9\%-confidence interval (corrected for multiple comparisons, corresponds to an $\alpha$-threshold of .001), 99.9UL: upper limit of the 99.9\%-confidence interval.} \\
\end{sidewaystable}

\begin{sidewaystable}
\caption{Correlations of well-being and productivity with all predictors at time 2}
\label{tab:cor2_complete}
\resizebox{17cm}{!}{
\begin{tabular}{@{}lSS[table-format=1.3]SSSSSS[table-format=1.3]SSSSSS[table-format=1.3]SSSSSS[table-format=1.3]SSSS@{}}
\toprule
                           & \multicolumn{6}{l}{\textbf{Well-being (Pearson’s r)}}                                                         & \multicolumn{6}{l}{\textbf{Well-being (Spearman’s rho)}}                                                         & \multicolumn{6}{l}{\textbf{Productivity (Pearson’s r)}}                                                       & \multicolumn{6}{l}{\textbf{Productivity (Spearman’s rho)}}                                                       \\
                           & \textit{\textbf{r}} & \textit{\textbf{p}} & \textbf{95LL} & \textbf{95UL} & \textbf{99.9LL} & \textbf{99.9UL} & \textit{\textbf{r\_s}} & \textit{\textbf{p}} & \textbf{95LL} & \textbf{95UL} & \textbf{99.9LL} & \textbf{99.9UL} & \textit{\textbf{r}} & \textit{\textbf{p}} & \textbf{95LL} & \textbf{95UL} & \textbf{99.9LL} & \textbf{99.9UL} & \textit{\textbf{r\_s}} & \textit{\textbf{p}} & \textbf{95LL} & \textbf{95UL} & \textbf{99.9LL} & \textbf{99.9UL} \\ \midrule
Well-being                 & 1                   & {-}                 & {-}           & {-}           & {-}             & {-}             & 1                      & {-}                 & {-}           & {-}           & {-}             & {-}             & .203                & .006                & .059          & .338          & -.040            & .423            & .273                   & 0                   & .133          & .402          & .035            & .482            \\
Productivity               & .203                & .006                & .059          & .338          & -.040            & .423            & .273                   & 0                   & .133          & .402          & .035            & .482            & 1                   & {-}                 & {-}           & {-}           & {-}             & {-}             & 1                      & {-}                 & {-}           & {-}           & {-}             & {-}             \\
Boredom                    & -.331               & 0                   & -.454         & -.195         & -.529           & -.099           & -.335                  & 0                   & -.457         & -.200           & -.532           & -.103           & -.145               & .050                 & -.284         & 0             & -.373           & .099            & -.172                  & .020                 & -.309         & -.027         & -.396           & .072            \\
Behavioral-disengagement   & -.408               & 0                   & -.522         & -.280          & -.590            & -.187           & -.404                  & 0                   & -.518         & -.275         & -.587           & -.182           & -.078               & .294                & -.221         & .068          & -.313           & .166            & -.111                  & .136                & -.252         & .035          & -.342           & .133            \\
Self-blame                 & -.396               & 0                   & -.511         & -.267         & -.581           & -.173           & -.381                  & 0                   & -.498         & -.25          & -.569           & -.156           & -.067               & .370                 & -.210          & .079          & -.302           & .177            & -.129                  & .082                & -.269         & .017          & -.358           & .115            \\
Need for Relatedness       & .477                & 0                   & .357          & .582          & .268            & .644            & .463                   & 0                   & .341          & .569          & .250             & .632            & .048                & .523                & -.098         & .191          & -.195           & .285            & .135                   & .069                & -.011         & .274          & -.109           & .363            \\
Need for Competence        & .375                & 0                   & .244          & .493          & .149            & .564            & .365                   & 0                   & .233          & .484          & .137            & .556            & .216                & .003                & .073          & .350           & -.026           & .434            & .255                   & 0                   & .114          & .386          & .015            & .467            \\
Need for Autonomy          & .543                & 0                   & .432          & .637          & .348            & .692            & .546                   & 0                   & .436          & .640           & .353            & .695            & .095                & .202                & -.051         & .237          & -.149           & .328            & .139                   & .060                 & -.006         & .279          & -.105           & .367            \\
Communication              & .386                & 0                   & .254          & .504          & .158            & .575            & .374                   & 0                   & .241          & .494          & .144            & .566            & .187                & .012                & .041          & .325          & -.059           & .412            & .252                   & .001                & .109          & .385          & .009            & .467            \\
Stress                     & -.540                & 0                   & -.635         & -.429         & -.691           & -.345           & -.522                  & 0                   & -.62          & -.408         & -.677           & -.323           & -.081               & .273                & -.224         & .064          & -.316           & .162            & -.092                  & .217                & -.234         & .054          & -.325           & .152            \\
Daily routines             & .425                & 0                   & .298          & .536          & .206            & .603            & .417                   & 0                   & .290           & .530           & .197            & .597            & .111                & .136                & -.035         & .252          & -.134           & .342            & .162                   & .028                & .018          & .300            & -.081           & .388            \\
Distractions               & -.330                & 0                   & -.453         & -.195         & -.528           & -.098           & -.342                  & 0                   & -.464         & -.208         & -.538           & -.112           & -.258               & 0                   & -.389         & -.118         & -.470            & -.019           & -.301                  & 0                   & -.427         & -.163         & -.505           & -.065           \\
Generalized anxiety        & -.526               & 0                   & -.623         & -.413         & -.680            & -.328           & -.541                  & 0                   & -.636         & -.430          & -.691           & -.346           & -.091               & .222                & -.233         & .055          & -.324           & .153            & -.124                  & .095                & -.264         & .022          & -.354           & .120            \\
Emotional loneliness       & -.449               & 0                   & -.557         & -.325         & -.622           & -.234           & -.477                  & 0                   & -.581         & -.356         & -.643           & -.267           & -.162               & .028                & -.300           & -.018         & -.388           & .081            & -.169                  & .022                & -.307         & -.025         & -.393           & .075            \\
Social loneliness          & -.469               & 0                   & -.575         & -.348         & -.637           & -.259           & -.480                   & 0                   & -.584         & -.361         & -.646           & -.272           & -.079               & .286                & -.222         & .067          & -.314           & .164            & -.118                  & .111                & -.259         & .027          & -.349           & .126            \\
Quality of social contacts & .533                & 0                   & .421          & .629          & .336            & .685            & .531                   & 0                   & .419          & .628          & .334            & .684            & .123                & .097                & -.023         & .263          & -.121           & .353            & .157                   & .034                & .012          & .295          & -.087           & .383            \\
Extraversion               & .276                & 0                   & .137          & .405          & .039            & .484            & .257                   & 0                   & .116          & .387          & .018            & .468            & .077                & .303                & -.069         & .219          & -.167           & .311            & .074                   & .319                & -.072         & .217          & -.169           & .309            \\
Quality of Sleep           & .481                & 0                   & .361          & .585          & .272            & .646            & .486                   & 0                   & .367          & .589          & .278            & .650             & .142                & .054                & -.003         & .282          & -.101           & .370             & .138                   & .062                & -.007         & .278          & -.105           & .367            \\
Age                        & -.171               & .020                 & -.308         & -.027         & -.394           & .072            & -.156                  & .035                & -.294         & -.011         & -.381           & .087            & .026                & .729                & -.120          & .170          & -.216           & .265            & -.008                  & .910                 & -.153         & .137          & -.248           & .233            \\ \bottomrule
\end{tabular}}

\resizebox{17cm}{!}{
\textit{Note}. \textit{r}: Pearson correlation coefficient, \textit{$r_s$}: Spearman's rank correlation coefficient (Spearman's rho), \textit{p}: $p-value$ (a $p-value$ of 0 indicates p < .001), 95LL: lower limit of the 95\%-confidence interval, 95UL: upper limit of the 95\%-confidence interval, 99.9LL: lower limit of the 99.9\%-confidence interval (corrected for multiple comparisons, corresponds to an $\alpha$-threshold of .001), 99.9UL: upper limit of the 99.9\%-confidence interval.} \\

\end{sidewaystable}

\end{document}